\documentclass[twocolumn,tighten,time]{aastex631}
\usepackage[fleqn]{amsmath}
\usepackage{graphicx}	
\pdfoutput=1 
\usepackage{epsfig}
\usepackage{subfigure}
\usepackage{multirow}

\newcommand{\der}{{\rm d}} 
 
\newcommand{\x}{^{\rm x}} \newcommand{\bb}{_{\rm m}} 
 \newcommand{\s}{_{\rm s}} 
 \newcommand{\m}{_{\rm m}}  
\newcommand{\R}{{R_{\rm f}}}  
 \newcommand{\cc}{_{\rm c}}

\newcommand{\pk}{_{\rm pk}} 
 \newcommand{\T}{{\Re}} 
 \newcommand{\h}{_{\rm h}}

\newcommand{\F}{^{\rm th}}  
\newcommand{\G}{{\cal G}} \newcommand{\p}{_{\rm up}}
\newcommand{\uu}{_{\rm u}} \newcommand{\f}{{\it f}} 
\newcommand{\up}{_{\rm up}}

\newcommand{\ti}{t_{\rm i}} \newcommand{\e}{_{\rm e}}

\newcommand{\beq}{\begin{equation}} \newcommand{\eeq}{\end{equation}}

 \newcommand{\beqa}{\begin{eqnarray}}
\newcommand{\eeqa}{\end{eqnarray}} \newcommand{\lav}{\langle}
\newcommand{\rav}{\rangle} 
\newcommand{\vir}{_{\rm vir}}

\begin{document}

\shorttitle{Halo Bias in the Peak Model}
\shortauthors{Salvador-Sol\'e \& Manrique}

\title{HALO BIAS IN THE PEAK MODEL. A FIRST-PRINCIPLES NONPARAMETRIC APPROACH}

\author{Eduard Salvador-Sol\'e and Alberto Manrique}
\affiliation{Institut de Ci\`encies del Cosmos. Universitat de Barcelona, E-08028 Barcelona, Spain}

\email{e.salvador@ub.edu}



\begin{abstract}
The Press-Schechter (PS) and excursion set (ES) models of structure formation fail in reproducing the halo bias found in simulations, while the excursion set-peaks (ESP) formalism built in the peak model reproduces it only at high masses and does not address in a fully satisfactory manner peak nesting and the mass and time of ellipsoidal collapse of triaxial peaks in the Gaussian-smoothed density field. Here we apply the CUSP formalism fixing all these issues from first principles and with no free parameters to infer the Lagrangian local peak bias parameters, which adopt very simple analytic expressions similar to those found in the PS and ES models. The predicted Eulerian linear halo bias recovers the results of simulations. More specifically, we show that the only small departure observed at intermediate and low masses can be due to the spurious halo splitting and grouping caused by the Spherical Overdensity halo-finding algorithm used in simulations. 
\end{abstract}



\section{INTRODUCTION}\label{intro}

Galaxies are biased tracers of matter. Determining this bias is very challenging as it involves not only the galaxy distribution in dark matter (DM) halos (e.g. \citealt{CS02,Bea02,Yea03,Zeha05,Zea05}), but also the biased distribution of halos themselves (e.g. \citealt{HP73,BS83}). 

Simulations show that the two-point correlation function of halos of mass $M$ is nearly proportional to the matter correlation function in the density field smoothed on that mass scale, $\xi\h(r)\sim b^2\xi\bb(r)$, where $b$ is an increasing function of $M$, i.e. the more massive halos, the more clustered (e.g. \citealt{MW96,ST99,SMT01}, hereafter SMT; \citealt{Tea01,SW04,Tea05,Tea10}, hereafter T+10; \citealt{Lea16}, hereafter LWBS). Since there is no interaction between halos able to yield their mass segregation, this `bias' must already affect protohalos in the linear density field and, hence, be imprinted in the power spectrum of density perturbations. 

A more precise relation between $\xi\h(r)$ and $\xi\bb(r)$ is given by the `perturbative bias expansion' of the former in powers of the latter. The most practical way to derive it theoretically is by inferring first the Lagrangian form holding for protohalos in the so-called `peak-background split' (PBS) approximation (\citealt{BBKS}, hereafter BBKS) and then the Eulerian form for halos (see the review by \citealt{Dea18}). 

At first order, the perturbative bias expansion is equivalent to the simple `linear' relation $\delta\h\approx b_1\delta\m$ between the protohalo overdensity $\delta\h$ and the matter density contrast $\delta_{\rm m}$ \citep{K84,Eea88}, i.e. the leading order of the alternative ``local-in-matter bias expansion'' (hereafter simply the local expansion) of $\delta\h$ in powers of $\delta_{\rm m}$ \citep{FG93}. 

\citet{CK89} derived for the first time the linear bias from the conditional mass function (MF) of halos lying on a background in the Press-Schechter (1974; PS) model of structure formation using top-hat smoothing. \citet{MW96} extended this derivation to the conditional halo MF in the `excursion set' (ES) model using a $k$-sharp (or $k$-space top-hat) filter in order to correct for cloud-in-cloud configurations \citep{BCEK}. SMT showed that the unconditional MF can be used as well, though \citet{Dea10} (hereafter DCSS) noticed that this procedure yields the loss of the scale-dependent bias component. 

But all these models assume top-hat spherical collapse, characterized by a mass-independent critical density contrast for collapse (or fixed barrier in the ES formalism), which leads to a deficient halo MF. \citet{ST02} introduced a parameterized mass-dependent density contrast for collapse (or moving barrier) so as to account for top-hat ellipsoidal collapse (see also \citealt{Jea01}). This greatly improved the predicted halo MF though not yet enough the linear bias. 

These results suggested that the PS/ES models do not correctly predict the halo bias because they do not yield a fine enough description of halo formation. Indeed, in these models, protohalos are fully characterized by their height $\nu$ at scale $R$ fixing the time $t$ of collapse of halos with mass $M$ regardless of their environment, so they cannot explain another result of simulations, that halos of a given mass lying on different backgrounds have distinct formation times and internal properties \citep{ST04,Gea05,Hea06,Wech06,Wet07,GW07,Mea18,Hea20}, the so-called `secondary (or assembly) halo bias'. Only if the filter was different than the $k$-sharp one (leading to correlated random walks) did the ES model predict halo formation times dependent on background \citep{Z07}. Certainly, the filter was undetermined and the resulting halo MF was not analytic, but this result showed the possibility of explaining the secondary bias provided a good enough model. 

\citet{Dea08} noted that in the more realistic peak model of structure formation where halos form from the collapse of patches around density maxima or peaks, protohalos are characterized, apart from the peak height $\nu$ and scale $R$, by the curvature $x$, which depends on the background. Consequently, the peak model could lead in a simple natural manner to the secondary halo bias, and likely also provide a better description of the primary bias. 

Unfortunately, the peak model faces several fundamental difficulties: i) the peak number density derived by BBKS is per infinitesimal height instead of per infinitesimal scale as needed to infer the halo MF \citep{MSS95,PSD13}; ii) peaks are characterized by the first and second derivatives of the density field making Gaussian smoothing compulsory, but the mass associated with a Gaussian peak is hard to tell; iii) peaks are triaxial and the time of ellipsoidal collapse is also unknown; and iv) the peak number density at $\ti$ must be corrected for peak nesting (the peak version of the cloud-in-cloud problem; \citealt{AJ90}). Yet, \citet{PS12} and \citet{PSD13} managed to build the `excursion set-peaks' (ESP) formalism, inspired in the ES model but monitoring the random walks of peaks instead of fixed points. ESP predicts a halo MF and local bias in unprecedented agreement with the results of simulations (\citealt{PSD13}; LWBS). However, ESP is not fully satisfactory. It does not count objects {\it first} crossing the barrier, but simply crossing it. This way it can use the Gaussian filter,\footnote{There is no need then to use a $k$-sharp filter to deal with {\it uncorrelated} walks.} necessary to deal with peaks, but this procedure is insufficient to correct for peak nesting. In addition, the mass and top-hat ellipsoidal collapse time of Gaussian peaks, chosen in a motivated way, are parametrized. 

But all the difficulties of the peak model are fixed in the {\it ConflUent System of Peak trajectories} (CUSP) formalism {\it from first principles and with no free parameter} \citep{SM19}. CUSP allows one, indeed, to infer the peak number density per infinitesimal scale $R$ corrected for peak nesting \citet{MSS95} and to unambiguously determine the mass and typical collapsing time of Gaussian peaks \citep{Jea14a}. CUSP reproduces the MF of simulated halos \citep{Jea14b} for any given halo mass definition, so it could also successfully predict the halo bias. In addition, CUSP makes the link between peaks with $\nu$ and the {\it average} curvature $x$ at $R$ and the average density profile \citep{Sea12a,Vea12}, concentration \citep{Sea23}, kinematics and shape \citep{Sea12b} as well as substructure \citep{I,II,III} of halos with $M$ at $t$, which are also well-predicted, so it stands as an ideal tool for trying to explain the secondary bias too. Here we address the former issue, while the latter is addressed in a forthcoming Paper. 

The layout of the present Paper is as follows. In Section \ref{CUSP} we remind how CUSP determines the mass and collapse time of Gaussian triaxial peaks. In Section \ref{unconditional} we derive the conditional peak number density per infinitesimal scale, which is corrected for nesting in Section \ref{nesting}. In Section \ref{LB} we derive the Lagrangian local peak bias parameters. The predicted Eulerian linear halo bias is compared to the results of simulations in Section \ref{sim}. In Section \ref{relax} we correct simulations for spurious halo splitting and grouping and repeat the comparison. The results are summarized and the main conclusions are drawn in Section \ref{dis}.

\section{Peak Mass and Ellipsoidal Collapse Time}\label{CUSP}

The time of ellipsoidal collapse of triaxial patches around Gaussian peaks at $\ti$ depends not only on their mass and size like in top-hat spherical collapse, but also on their shape and concentration (e.g. \citealt{P80}). In other words, it is a function of the density contrast $\delta$, smoothing radius $R$, ellipticity $e$, prolateness $p$, and curvature $x$ of the peaks. However, the probability distributions functions (PDFs) of $e$, $p$ and $x$ of peaks with $\delta$ at $R$ are very sharply peaked (BBKS), so all patches traced by peaks with given $\delta$ and $R$ have very similar values of $e$, $p$ and $x$ and, hence, collapse at essentially the same time. In other words, the typical time $t$ of Gaussian ellipsoidal collapse (neglecting the small scatter or stochasticity) of such patches essentially depends on $\delta$ and $R$, like in top-hat spherical collapse. Consequently, for any given $\delta(t)$ relation, we can find the radius $R$ of the Gaussian filter such that the collapsing patches at $\ti$ traced by peaks with $\delta$ at $R$ give rise to halos with mass $M$ at $t$. Thus, those $\delta(t)$ and $R(M,t)$ relations establish, by construction, a one-to-one correspondence between halos with $M$ at $t$ and peaks with $\delta$ on $R$ at $\ti$. We remark that such a halo-peak correspondence holds regardless of whether the collapse is monolithic (the halo grows by pure accretion) or lumpy (it undergoes major mergers). 

As shown by \citet{Jea14a}, the $\delta(t)$ and $R(M,t)$ relations are fully determined by the consistency conditions that: i) all the DM in the Universe at any $t$ is locked inside halos, and ii) the mass $M$ of halos is equal to the volume-integral of their density profile. Specifically, if we write the density contrast $\delta$ and the rms density fluctuation (or 0th-order spectral moment) $\sigma_0$ of Gaussian peaks of scale $R$ collapsing ellipsoidally into halos with $M$ at $t$ as proportional to the homologous quantities in top-hat smoothing (all quantities referring to top-hat smoothing are hereafter denoted with index `th' to distinguish them from those referring to Gaussian smoothing), 
\beq
\delta(t,\ti)=r_\delta(t)\,\delta\F(t,\ti)
\label{deltat}
\eeq
with $\delta\F(t,\ti)=\delta\cc\F(t)D(\ti)/D(t)$, where $\delta\F\cc(t)$ is the critical linearly extrapolated density contrast for spherical collapse at $t$ (equal to 1.686 in the Einstein-de Sitter universe) and $D(t)$ is the linear growth factor, and
\beq
\sigma_0(R,\ti)=r_\sigma(M,t)\,\sigma_0\F(M,\ti)\,.
\label{rm}
\eeq
The function $r_\delta(t)$ so obtained depends on cosmology and the function $R(M,t)$ depends in addition on the halo mass definition adopted (i.e. their assumed mean overdensity; see below). In all cases analyzed (see below), they appear to be reasonably well fitted by the simple analytic expressions, 
\beqa 
r_\delta(t)\approx \frac{a^{d{\cal D}(t)}(t)}{D(t)}
\,~~~~~~~~~~~~~~~~~~~~~~~~~~~~~~~~~~
\nonumber\\
{\cal D}(t)=1-d_0a^{0.435/a(t)}(t)\phantom{\frac{}{}}~~~~~~~~~~~~~~~~~~~~~~
\label{cc}\\
r_\sigma(M,t)\approx 1+r_\delta(t){\cal S}(t)\nu\F(M,t)~~~~~~~~~~~~~
\nonumber\\
{\cal S}(t)=s_0\!+\!s_1a(t)\!+\!\log\left[\frac{a^{s_2}(t)}{1\!+\!a(t)/A}\right],~~~~~~~~
\label{rs}
\eeqa
with $\nu\F(M,t)\equiv\delta\F(t,\ti)/\sigma\F_0(M,\ti)=\delta\F\cc(t)/\sigma\F_0(M,t)$ equal to the height in top-hat smoothing. Equation (\ref{cc}) is a refinement at low-$z$ of the homologous expression given in previous works (see App.~\ref{C}). According to equation (\ref{deltat}), $\delta\cc=r_\delta(t)\delta\cc\F(t)$ is the linearly extrapolated critical density contrast for Gaussian ellipsoidal collapse at $t$. For simplicity in the notation, we will skip from now on the explicit dependence of $\delta$ and $\sigma_0$ on the (arbitrary) value of $\ti$.

\begin{figure}
\centerline{\includegraphics[scale=1.05, bb= -125 20 400 258]{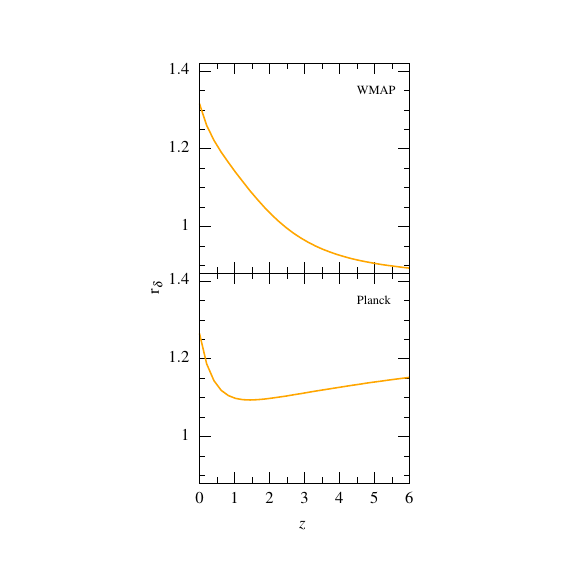}}
 \caption{Factor $r_\delta$ as a function of redshift for the cosmologies given in Table 1.}
 \label{f0}
\end{figure}

\begin{figure}
\centerline{\includegraphics[scale=1.05, bb= -125 20 400 258]{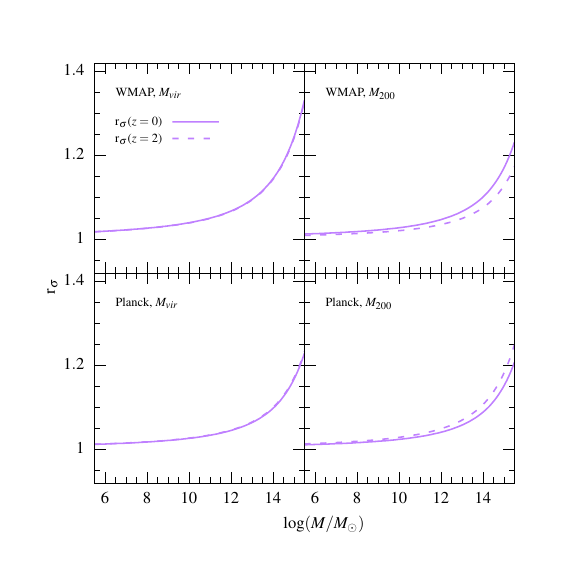}}
 \caption{Factor $r_\sigma$ as a function of mass at $z=0$ and $z=2$ for the halo mass definitions and cosmologies given in Table 1. The curves $r_\sigma$ for $M\vir$ masses at $z=0$ and $z=2$ overlap in all cosmologies.}
 \label{f1}
\end{figure}

In Table \ref{T1} we provide the values of coefficients $d$, $d_0$, $s_0$, $s_1$, $s_2$ and $A$ for some mass definitions and cosmologies of interest, specifically for $M_{200}$ and $M_{\rm vir}$ masses defined in the Table caption, and the {\it WMAP7} and {\it Planck14} cosmologies given in Table \ref{T2}. To illustrate how the functions $r_\delta(t)$ and $r_\sigma(M,t)$ depend on cosmology and mass definition, we plot them in Figures \ref{f0} and \ref{f1} for the particular cases quoted in those Tables.  

\begin{table}
\caption{Coefficients in the halo-peak relations.}
\begin{center}
\begin{tabular}{ccccccccccc}
\hline \hline
Cosmol.\! & Mass$\!$ & $d$ & $10d_0\!\!$ & $10^{2}s_0\!\!$ & $10^{2}s_1\!\!$ & $10^{2}s_2\!\!$ & $A$ \\ 
\hline
\multirow{2}{*}
{WMAP7}\! & \!Mass$^*\!$ & 1.06 & 3.0 & 4.22 & 3.75 & 3.18 & 25.7\\ 
   & $M_{200}\!$ & 1.06 & 3.0 & 1.48 & 6.30 & 1.32 & 12.4\\ 
\multirow{2}{*}
{Planck14}\! & $M\vir\!$ & 0.93 & 0.0 &  2.26 & 6.10 & 1.56 & 11.7 \\ 
   & $M_{200}\!$ &  0.93 & 0.0 & 3.41 & 6.84 & 2.39 & 6.87 \\
\hline
\end{tabular}
\end{center}
\label{T1}
$^*M\vir$ and $M_{200}$ are the masses inside the region with a mean inner density equal to $\Delta\vir(t)$ \citep{bn98} times the mean cosmic density, and 200 times the critical cosmic density, respectively.\\
\end{table}

\begin{table}
\caption{Cosmological parameters.}
\begin{center}
\begin{tabular}{ccccccc}
\hline \hline 
Cosmology& $\Omega_\Lambda$ & $\Omega_{\rm m}$ & $h$ &
$n_{\rm s}$ & $\sigma_8$ & $\Omega_b$\\ 
\hline 
WMAP7$^{a}$ & 0.73 & 0.27 & 0.70 & 0.95 & 0.81 & 0.045\\
Planck14$^{b}$ & 0.68 & 0.32 & 0.67 & 0.96 & 0.83 & 0.049\\ 
\hline
\end{tabular}
\label{T2}
\end{center}
$^{a}$ \citet{Kea11}.\\
$^{b}$ \citet{P14}.
\end{table}

\begin{figure}
\centerline{\includegraphics[scale=1.05, bb= -125 20 400 258]{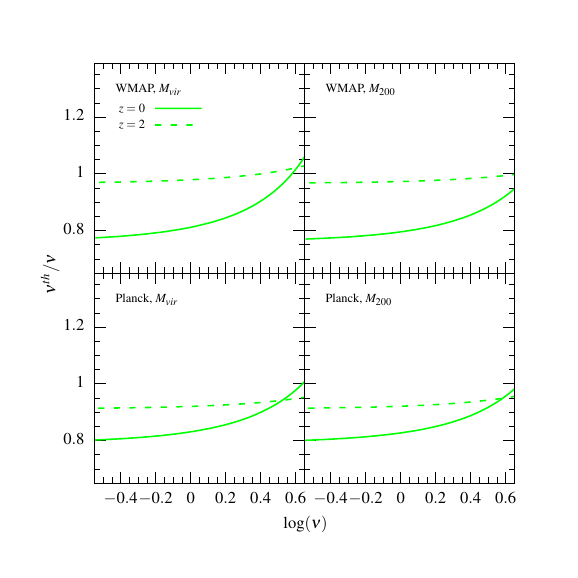}}
 \caption{Comparison between the peak height in top-hat smoothing and spherical collapse and Gaussian smoothing and ellipsoidal collapse for the same cases as in Figure 1. Again, we plot the results for $z=0$ and $z=2$.}
 \label{f2}
\end{figure}

Note that the scale $R$ of peaks with $\delta(t)$ leading to halos with $M$ at $t$, given by the implicit equation (\ref{rm}), is a function of both arguments, $R(M,t)$, contrarily to what happens in the PS/ES models. An explicit expression for this function could be obtained if, instead of equation (\ref{rm}), we had adopted the proportionality $R(M,t,\ti)\equiv r_{\rm R}(M,t)\,R\F(M,\ti)$, with $R\F=[3M/(4\pi\bar\rho(\ti))]^{1/3}$ where $\bar\rho$ is the mean cosmic density, and determined the proportionality factor $r_R(M,t)$ as done with $r_\delta$ and $r_\sigma$. However, such an explicit expression for $R(M,t)$ cannot be adopted simultaneously to that of $\sigma_0(R)$ (eq.~[\ref{rm}]) and we prefer to use this latter since it allows one to directly relate the time-invariant peak height in Gaussian smoothing and ellipsoidal collapse to that in top-hat smoothing and spherical collapse, through the simple relation 
\beq
\nu(M,t)\approx \frac{r_\delta(t)}{r_\sigma[R(M,t),t]}\nu\F(M,t).
\label{nus}
\eeq
This relation is also shown in Figure \ref{f2} at two different redshifts for the different cases gathered in Table 1.

As mentioned, the scale $R$ of peaks is, in general, a function of $M$ and $t$ (or $\delta$) of halos and so is also $\sigma_0$ as well as any higher order spectral moment $\sigma_j$, related in Gaussian smoothing to $\sigma_0$ through the recursive relation 
\beq
\frac{\der \sigma_j^2}{\der R^2}=-\sigma_{j+1}^2.
\label{j}
\eeq
There is, however, one exception. For the reason explained in \citet{Jea14b}, when using $M\vir$ masses, $R$ turns out to be a function of $M$ alone (and vice versa), as seen in Figure \ref{f1}, where for $M\vir$ masses, $r_\sigma$ and, hence, $\sigma_0$ is independent of $z$. Equation (\ref{rs}) then implies $r_\delta(t)= c[{\cal S}(t)\delta\F(t,\ti)]^{-1}$ and $r_\sigma= 1+c/\sigma_0\F(M,\ti)$, where the constant $c$ is equal to 0.14 and 0.10 in the {\it WMAP7} and {\it Planck14} cosmologies, respectively. 

We emphasize that all the previous expressions and the values of their coefficients do not respond to any parametrization of the model and its tuning against numerical simulations. As explained, they are simple analytic fits to the (accurate) numerical relations implied by self-consistency conditions of the model. In this sense, the mass and collapse time of Gaussian peaks are found in CUSP from first principles with no free parameter.    

\section{Peak Number Density per Infinitesimal Scale}\label{unconditional}

We are now ready to calculate the conditional peak number density needed to derive the linear halo bias in the PBS approach. This latter derivation will be achieved in Section \ref{LB} taking into account the relations $\delta(t)$ and $R(M,t)$ (or $R(M,\delta)$) passing from peaks to halos. According to these relations, the scale $R$ of the peak giving rise to a halo with $M$ at $t$ depends (in general) not only on $M$ but also on $t$ or, equivalently, on $\delta$, so from this point of view $R$ and $\delta$ are not independent arguments. However, in the present Section we are only concerned with peaks in the smoothed density field at $\ti$ regardless of the peak-halo correspondence, so there is no relation between the $R$ and $\delta$ arguments. 

As shown in \citet{MSS95}, the density contrast of Gaussian peaks decreases monotonically with increasing scale. Consequently, peaks crossing $\delta$ at scales between $R$ and $R+ \der R$ are those with density contrast $\tilde\delta$ greater than $\delta$ at scale $R$ that have density contrast smaller than $\delta$ at $R + \der R$. Thus, taking into account the general relation for Gaussian smoothing, 
\beq 
\frac{\partial\delta({\bf r},R)}{\partial R}= R \nabla^2\delta({\bf r},R)\equiv -x({\bf r},R)\sigma_2(R)R\,
\label{0th}
\eeq
$\tilde\delta$ must satisfy the condition
\beq
\delta < \tilde\delta \le \delta+\sigma_2(R)\,\tilde x\,R\,\der R\,,
\label{cond}
\eeq 
where $\tilde x$ is their curvature at $R$. Consequently, the average {\it unconditional} number density $N\pk(R,\delta)$ of peaks with $\delta$ at scales between $R$ and $R+\der R$ is the integral of the average number density of peaks per infinitesimal height (or density contrast) and curvature, derived by BBKS, over $\tilde \delta$ in the range delimited by the inequality (\ref{cond}) over $\tilde x$ in the whole range of (positive) values. This way we obtain
\beq
N\pk(R,\delta)
=\frac{G_1(R,\delta)}{(2\pi)^2\,R_{\star}^3}\,
{\rm e}^{-\frac{\nu^2}{2}}\,
\frac{\sigma_2}{\sigma_0}
\,R\,,
\label{npeak}
\eeq
where $R_\ast=\sqrt{3}\,\sigma_1/\sigma_2$, 
\beqa
G_i(R,\delta)=\int_0^\infty\! \der x\,\frac{x^i F(x) {\rm e}^{-\frac{(x-x_\ast)^2}{2(1-\gamma^2)}}}{\sqrt{2\pi(1-\gamma^2)}},~~~~~~~~~~~~~~~~~~\\
F(x)\!=\!\frac{(x^3-x)\left\{{\rm erf}\left[\left(\frac{5}{2}\right)^{\frac{1}{2}} x\right]\!+{\rm erf}\left[\left(\frac{5}{2}\right)^{\frac{1}{2}}\frac{x}{2}\right]\right\}}{2\!+\!\left(\frac{2}{5\pi}\right)^{\frac{1}{2}}\!\!\left[\!\left(\frac{31x^2}{4}+\frac{8}{5}\right){\rm e}^{-\frac{5x^2}{8}}\!\!+\!\left(\frac{x^2}{2}-\frac{8}{5}\right)\!
{\rm e}^{-\frac{5x^2}{2}}\right]},~~~~\nonumber
\eeqa
$\gamma\equiv\sigma_1^2/(\sigma_0\sigma_2)$, and $x_\ast\equiv \gamma \nu$. Alternatively, function $G_1(R,\delta)$ in equation (\ref{npeak}) can be written in terms of the average curvature of peaks with $\delta$ at $R$ through the relation (BBKS)
\beq
\lav x\rav(R,\delta)=\!\frac{G_1(R,\delta)}{G_0(R,\delta)}
\label{hat}
\eeq
(hereafter we skip the arguments of $\lav x\rav$ and of $\lav x^2\rav=G_2/G_0$ used below) so that $N\pk(R,\delta)$ is $\lav x\rav R \sigma_2/\sigma_0$ times the number density of peaks with $R$ per infinitesimal $\nu$, ${\cal N}(\nu,R)$, calculated by BBKS.\footnote{Factor $G_0$ in ${\cal N}(\nu,R)$ is missing in equations (14) and (15) of \citet{Jea14b}, though it is included in the calculations, so all the results reported in that work are correct.} This peak number density leads, through the $\delta(t)$ and $M(R,t)$ relations, to the unconditional halo MF shown in \citet{Jea14b} to reproduce the results of simulations.

But what we actually need to derive the halo bias is the average {\it conditional} number density of peaks per infinitesimal scale constrained to lie on a background. Repeating the same procedure above, starting from the conditional number density of peaks per infinitesimal $\nu$ at $R$ subject to having $\nu'$ at $R'$ calculated by BBKS, we obtain
\beq 
N\pk(R,\delta|R',\delta')
=\frac{G_1(R,\delta|R',\delta')}{(2\pi)^2\,R_{\star}^3\,e}
      \,{\rm e}^{-\frac{\left(\nu-
            \epsilon\nu''\right)^2}{2e^2}}\,{\frac{\sigma_2}{\sigma_0}}R,~~~~
\label{mf16}
\eeq 
where $\nu''=\delta'/\sigma_0'$, $\gamma'=(\sigma_1')^2/(\sigma_0'\sigma_2')$, with a prime on $\sigma_j'$ denoting that it is a function of $R'$, $\epsilon=\sigma_0^2(R_{\rm h})/[\sigma_0\sigma_0']$, and $e=\sqrt{1 - \epsilon^2}$, with $R_{\rm h}=[(R^{2}+R'^{2})/2]^{1/2}$. In equation (\ref{mf16}), $G_1(R,\delta|R',\delta')$ is defined as the unconditional $G_1$ but with $x_\ast=\gamma\nu$ replaced by $\tilde x_\ast=\tilde\gamma\tilde\nu$ \citep{MSS95}, being
\begin{equation}
\tilde\gamma=\gamma\left[1+\epsilon^{2}\frac{(1-r_{1})^{2}}{(1-\epsilon^{2})}\right]^{1/2}
\label{2}
\end{equation}
\begin{equation}
\tilde\nu=\frac{\gamma}{\tilde\gamma}\frac{(1-r_{1})}{(1-\epsilon^{2})}\left[\nu\frac{(1-\epsilon^{2}r_{1})}{(1-r_{1})}-\epsilon\nu''\right],
\label{3}
\end{equation}
with $r_{1}=\sigma^2_{\rm 1h}\sigma^{2}_{0}/\sigma^{2}_{\rm 0h}\sigma^{2}_{1}$ and $\sigma_{jh}$ defined as $\sigma_{j}$ but with $R$ replaced by the $R_{\rm h}$. The same change affects all $G_i$ for conditional peaks, so it also affects the average conditional curvature $\lav x\rav/R,\delta|R',\delta')$ according to the definition (\ref{hat}).

Notice that, in the limit $R'\rightarrow \infty$, the conditional peak number density becomes the unconditional one, i.e. with no reference to the background density contrast $\delta'$ which vanishes. This makes a big difference with respect to what happens in the ES model where the limit $R'\rightarrow \infty$ in the conditional MF of halos does not lead to the unconditional MF \citep{MW96}. This difference will have important consequences in Section \ref{LB}.

Interestingly, at $R'> 3R$, we can neglect $R^2/(R')^2$ in front of unity, which leads to
\beqa
\epsilon\approx\frac{\sigma_0(R'/\sqrt{2})}{\sigma_0(R)} \frac{\sigma_0(R'/\sqrt{2},)}{\sigma_0(R',\delta)}\,~~~~~~~~~~~~~~~~~~~~~~\label{firstbis}\\
r_1\approx  \frac{\sigma_1^2(R'/\sqrt{2})}{\sigma_1^2(R)} \frac{\sigma_0^2(R)}{\sigma_0^2(R'/\sqrt{2})}.~~~~~~~~~~~~~~~~~~~~~\label{second}
\eeqa
Then, $r_1$, equal to $(\sqrt{2}R/R')^{2}$ in power-law spectra and close to it in the CDM spectrum, can also be neglected in front of unity. And so can also $\epsilon^2$ as $\sigma_0^2(R'/\sqrt{2})/\sigma_0^2(R)$ behaves as $(\sqrt{2}R/R')^{n+3}$, while $\sigma_0^2(R'/\sqrt{2})/\sigma_0^2(R')$ stays constant. Yet, the quantity
\beq
\epsilon\nu''\approx\left[\frac{\sigma_0(R'/\sqrt{2})}{\sigma_0(R')}\right]^2\nu'
\label{third}
\eeq
(not to mix up $\nu'=\delta'/\sigma_0(R)$ with $\nu''=\delta'/\sigma_0(R')$) is not negligible, so the conditional peak number density resulting from this approximation differs from the unconditional one (see eqs.~[\ref{2}] and [\ref{3}]). Therefore, at $R'>3 R$, we have $\tilde\gamma\approx\gamma$ and $\tilde\nu\approx (\nu-\epsilon\nu')$ (see eqs.~[\ref{2}]-[\ref{3}]), implying
\beqa
\tilde \gamma\tilde \nu\approx  \gamma
[\nu-q(R')\nu']~~~~~~~~~~~~~~~~~~~~~~~~~~~~~~~~~\label{one}\\
\frac{\nu-\epsilon\nu'}{e}\approx \nu-q(R')\nu',~~~~~~~~~~~~~~~~~~~~~~~~~~~~~\label{two}
\eeqa
where $q(R')=[\sigma_0(R'/\sqrt{2})/\sigma_0(R')]^2$. For power-law power spectra, $P(k)=Ck^n$, $q$ is constant and equal to $2^{(n+3)/2}$, while for the Cold Dark Matter (CDM) spectrum it is also roughly so in the relevant mass range. Thus, the message to take away is that $q(R')$ is a little sensitive to the value of $R'$; this is why it is hereafter written simply as $q$.

Taking into account the relations (\ref{one}) and (\ref{two}) holding for $R'> 3 R$, $N\pk(R,\delta|R',\delta')$ adopts the form of the unconditional number density of peaks with density contrast $\delta-q\delta'$, $N\pk(R,\delta\!-\!q\delta')$, almost independent of the background scale $R'$.\footnote{The same is true for the average curvature of conditional peaks. This will be used in the forthcoming Paper.}  This result greatly resembles that met in the PS model where the conditional number density of protohalos with $\delta\F$ at $R\F$ lying on a background $(\delta')\F$ coincides with the unconditional number density of protohalos with density contrast $\delta\F-(\delta')\F$ regardless of the background scale $(R')\F$ \citep{Eea88}. 

\section{Correction for Nesting}\label{nesting}

But the previous unconditional and conditional number densities refer to {\it all} peaks, while some peaks are nested within other larger scale peaks with the same density contrast, implying that they will be captured by more massive halos before completing their collapse and give rise to subhalos \citep{I} rather than halos. Therefore, the previous conditional peak number density needs to be corrected for peak nesting. In this respect, it is interesting to realize that the preceding derivation of the conditional peak number density per infinitesimal scale is nothing but the `one-step' barrier crossing procedure used in ESP (with the density contrast for Gaussian instead of top-hat ellipsoidal collapse), so the conditional peak number density in ESP is actually not corrected for nesting. This would explain why the linear bias it predicts is deficient at low masses ($\log\nu\F\la 0.0$) where peak nesting is significant (see below). 

The peak nesting correction factor we are going to infer is the same for both the conditional and unconditional number densities, so we will focus on the latter as it greatly simplifies the notation. The unconditional number density of peaks per infinitesimal scale {\it corrected for nesting}, $n\pk(R,\delta)$, satisfies the relation (\citealt{MSS95})
\beq
n\pk(R,\delta)
\approx N\pk(R,\delta)
-\!\int_R^\infty\! dR'\,N\pk(R,\delta|R',\delta)\, f(R',\delta),~
\label{nnp}
\eeq
where $f(R,\delta)=M(R,\delta)/\bar\rho(\ti) n\pk(R,\delta)$ is the mass fraction of peaks with $\delta$ per infinitesimal scale around $R$ at $\ti$. Equation (\ref{nnp}) is approximate because it uses the conditional number density of peaks lying {\it on a background}, instead of on a peak. This is why the mass fraction derived from it matches the results of simulations only at the 5\% level \citep{Jea14b}. A more accurate expression could be used \citep{Jea14b}, but that would complicate the calculations. The error this yields in the linear bias will be analyzed in Section \ref{LB}. 

\begin{table}
\caption{Best values of the coefficients entering the analytic correction for peak-nesting (with $q\e=2.4$).}
\begin{center}
\begin{tabular}{cccccc}
\hline \hline 
\!Cosmol.\!\! & Mass\!\! & $c_{00}\!$ & $c_{01}\!$ & $c_{10}\!$ & $c_{11}\!$ \\ 
\hline 
   \multirow{2}{*} 
   {WMAP7} 
   & \!$M\vir\!\!$ & $15.4\!$ & $-5.94\!$ & $21.7\!$ & $-7.83$ \\
   & \!$M_{200}\!\!$ & $21.9\!$ & $-9.34\!$ & $30.0\!$ & $-12.4 \!$\\
   \multirow{2}{*} 
   {Planck14} & $M\vir\!$ & $12.6\!$ & $-4.92\!$ & $17.7\!$ & $-6.62\!$ \\ 
   & \!$M_{200}\!\!$ & $11.0\!$ & $-4.31\!$ & $15.3\!$ & $-5.74\!$ \\
\hline
\end{tabular}
\label{T3}
\end{center}
\end{table}

\begin{figure}
\includegraphics[scale=1.2, bb= 35 10 400 198]{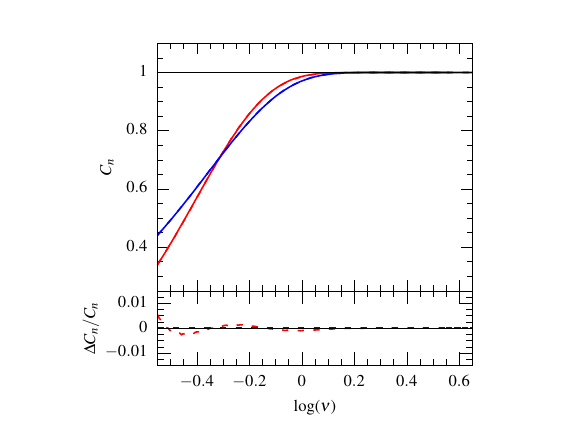}
 \caption{{\it Upper panel:} Analytic approximation of the peak-nesting correction factor (dashed line) in the {\it WMAP7} cosmology with $M_{\rm vir}$ masses at $z=0$ (red lines) and $z=2$ (blue lines) compared to its accurate numerical counterpart (solid line). {\it Lower panel}: Relative errors with respect to the accurate numerical solutions.}
\label{f3}
\end{figure}

\begin{figure}
\includegraphics[scale=1.2, bb= 35 10 400 198]{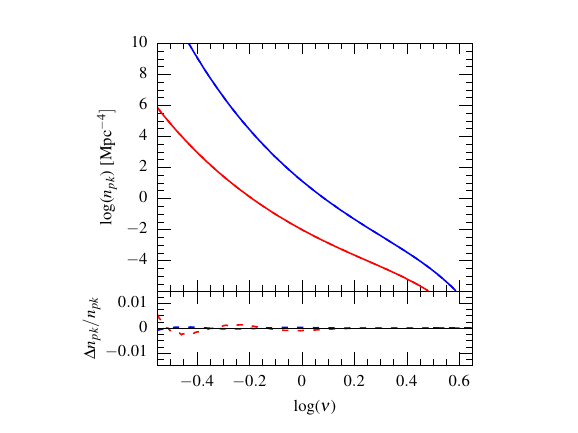}
 \caption{{\it Upper panel:} Same as Figure \ref{f3} but for the number density of non-nested peaks.}
\label{f4}
\end{figure}

The number density $n\pk(R,\delta)$ corrected for nesting is the solution of the integral equation (\ref{nnp}) of the Volterra type. But we are interested in deriving an analytic expression for the linear bias, so the numerical solution found in \citet{Jea14b} is not what we need here. Fortunately, an accurate enough analytic solution can also be found. Taking into account that $N\pk(R,\delta|R',\delta)$ can be approximated at $R'>3R$ by $N\pk(R,\bar q\delta)$, with $\bar q\equiv 1-q$ little dependent on $R'$, the nesting correction factor, $C_{\rm n}(R,\delta)\equiv n\pk(R,\delta)/N\pk(R,\delta)$ adopts the following approximate expression (see App.~\ref{A}) 
\beq
C_{\rm n}(R,\delta)\approx C(R,\delta)\!\left[1\!-\widetilde C(R,\delta) \right],
\label{ap}
\eeq
where
\beqa
C(R,\delta)= \frac{1-H(R,\delta)}{1-F\up(R,\delta)}~~~~~~~~~~~~~~~~~~~~~~~~~~~~~~~~\\
\widetilde C(R,\delta)\!=\!\frac{[c_1(\delta)\!-\!c_0(\delta)]H(R,\delta)}{1-H(R,\delta)}
\!\!\int_0^R\!\! \der R' C(R',\delta)f\uu(R',\delta).\nonumber\\
\label{ap2}
\eeqa
In equations (\ref{ap})-(\ref{ap2}), we have introduced the functions $f\uu\equiv M/\bar\rho(\ti)N\pk$, $H(R,\delta)\equiv N\pk(R,\bar\delta)/N\pk(R,\delta)$ and $F\up(R,\delta)\equiv c_1(\delta) R M(R,\delta)N\pk(R,\bar\delta)/\bar\rho(\ti)$ and evaluated $\bar q$ at some effective background scale $R'=R\e$. As explained in Appendix \ref{A}, to do this one must choose for any given $\delta$ one particular value of $q\e\equiv q(R\e)$ and obtain, by $\chi^2$-minimization of the difference between the analytic and numerical correction factors $C_{\rm n}$, the best values of coefficients $c_{i}(\delta)$. In all cases (i.e. cosmologies and mass definitions) and redshifts considered in this Paper, the minimum $\chi^2$ value is obtained for $q\e=2.4$, being the corresponding best fitting values of $c_i(\delta)$ very well fitted by simple linear relations  $c_i(\delta)=c_{i0}+c_{i1}\delta$, with the values of coefficients $c_{i0}$ and $c_{i1}$ depending on cosmology and halo mass definition (see Table \ref{T3}). 

As illustrated in Figures \ref{f3} and \ref{f4}, the approximate analytic correction factor $C_{\rm n}$ and the corrected peak number density, 
\beq
n\pk(R,\delta)=N\pk(R,\delta)C_{\rm n}(R,\delta),
\label{NCn}
\eeq
recover with high accuracy the respective numerical quantities down to a very low halo mass for all cases considered in this Paper. In particular, the solutions shown in these Figures, at $z=0$ and $z=2$ in the {\it WMAP7} cosmology and for virial masses ($M\vir$) are  accurate to better than $\sim 0.2$\% above $\log \nu\sim -0.45$. 

We emphasize that the coefficients appearing in $C_{\rm n}(R,\delta)$, like the coefficients in functions $r_\delta$ and $r_\sigma$ (eqs.~[\ref{deltat}] and [\ref{rm}]), have been obtained through the fit to the {\it accurate numerical predictions of CUSP}, not to the results of simulations. Therefore, this correction for nesting has been derived from first principles with no free parameter as the rest of CUSP.

\section{Local Bias Parameters}\label{LB}

The PBS approach was introduced by BBKS in the peak model by noting that the full density field smoothed on scale $R$ can be seen as the sum of the peak density field plus an independent background field smoothed on a larger scale $R'$. As shown by these authors, the local statistics of peaks in the latter composite case are nearly the same as for the full density field provided $R'> 3R$ and the matter correlation $\xi(r)$ of the background density field smoothed on scale $R'$ is nearly the same as that of the full density field smoothed on scale $R$ provided $r> 4R'$. 

In these conditions one can differentiate the {\it conditional} number density of peaks with $\delta$ at $R$ lying on a background $\delta\bb$ as done by DCSS, but with the following improvements: i) while the conditional peak number density used by DCSS was per infinitesimal $\nu$, ours is per infinitesimal $R$, ii) while the peak number density used by DCSS was not corrected for nesting, ours is, and iii) while DCSS find the Lagrangian linear peak bias only, we go further and derive the Eulerian linear halo bias by accurately accounting for the mass and time of ellipsoidal collapse of Gaussian peaks. In addition to these improvements described in previous Sections, we also apply an innovative procedure to address another important issue: the dependence of the conditional halo MF on the background scale.

In the PS model the conditional MF of halos lying on a background $\delta\bb$ does not depend on the background scale $R'$. As a consequence, the bias parameters are independent of $R'$. This is not the case in the ES model. Nevertheless, \citep{MW96} showed that it is possible to obtain scale-independent bias parameters of the PS kind by taking the limit $R'\rightarrow \infty$ in the conditional halo MF. The situation is even more complicated in the peak model because, as mentioned in Section \ref{unconditional}, taking the large-scale limit in the conditional peak number density, $N\pk(R,\delta|R',\delta\m)$, also dependent on $R'$ leads to the unconditional one, $N\pk(R,\delta)$, with no background at all. To circumvent this problem, DCSS took the limit $R'\rightarrow \infty$ in $\epsilon$ but not in $r_1$ (see eqs.~[\ref{mf16}]-[\ref{second}]), which is hard to justify. While in EPS this limit is taken in the cross-correlation of the conditional peak number density (normalized to the unconditional one) with Hermite polynomials of $\delta\m$ at the same time that the cross-correlation of $\sigma_0(R)$ and $\sigma_0(R')$ is set equal to one \citep{MPS12,PSD13}. Even though this procedure is mathematically consistent, its physical meaning is unclear. On the top of this, the large-scale limit in general is inconsistent with the condition $r>4R'$ for the  PBS approach to hold as it leaves no room for the separations $r$ in the correlations $\xi\h(r)$ and $\xi\m(r)$.

To avoid this problem, we will adopt an alternative procedure. Instead of taking the large-scale limit, we will use the result that, for $R'>3R$ as also required by the PBS approach, the conditional number density of non-nested peaks, $n\pk(R,\delta|R',\delta\bb)$, is essentially equal to $n\pk[R,\delta-q(R')\delta\bb]$. By adopting this approximation, the conditional number density of non-nested peaks depends on $R'$ only through the function $q(R')$, which, as mentioned, is essentially constant. Specifically, it is strictly equal to $2^{(n+3)/2}$ for power-law power spectra of index $n$ and about $1.6$ in the case of the CDM spectrum approximated by a power-law form of index $n\sim -1.7$ in the range of current background scales (or massive halos). Certainly, there is some uncertainty in that value of $q$, which depends on the exact range of background scales considered, but, as we will see below, any value of $q$ in the range $[1.65,1.55]$ makes a difference in the linear bias predicted with $q=1.6$ of less than 3\%, which is comparable to the statistical error in the results of simulations. Thus, the conditional peak number density for $R'>3R$ is very nearly given by $n\pk(R,\delta-q\delta\bb)$, i.e. it is essentially independent of $R'$ like in the ES model.

Then, Taylor expanding the overdensity of peaks lying on the background $\delta\bb$ for $\delta\bb$ around zero with that approximation for the conditional peak number density, we are led to
\beq
\!\delta\pk(R,\delta|\delta\bb)\!\equiv\!\frac{n\pk(R,\delta|\delta\bb)\!-\!n\pk(R,\delta)}{n\pk(R,\delta)}
\!\approx\!\sum_{i}\!\frac{B_i(R,\delta)}{i!}  \delta\m^i
\label{exp}
\eeq
\vspace{-18pt}
\beq
B_i(R,\delta)\!=\! \frac{1}{n\pk}\frac{\partial^{i} n\pk(R,\delta\!-\!q\delta\bb)}{\partial \delta\bb^i}\bigg|_{\delta\bb=0}
\!\!\!\!\!=\!\frac{(\!-q)^{i}}{n\pk}\frac{\partial^{i} n\pk(R,\delta)}{\partial \delta^i},
\label{B1}
\eeq
satisfying the recursive relation (with $B_0=1$)
\beq
B_i(R,\delta)
\!=\!-q B_{i-1}(R,\delta)\frac{\partial \ln(n\pk B_{i-1} )}{\partial \delta}.
\label{Bi}
\eeq
The local expansion (\ref{exp})-(\ref{B1}) is similar to that found in the ES/PS model. The parallelism is even stronger when writing it in the form
\beq
\!\delta\pk(R,\delta|\delta\bb)\!\approx\!\sum_{i}\frac{\widetilde B_i(R,\delta)}{i!}  (q\delta\m)^i
\label{exp2}
\eeq
\vspace{-18pt}
\beq
\widetilde B_i(R,\delta)\!\equiv\! \frac{1}{n\pk}\frac{\partial^i n\pk(R,\delta-q\delta\m)}{\partial (q\delta\m)^i}\bigg|_{q\delta\m=0}
\!\!\!=\frac{B_i(R,\delta)}{q^i}.
\label{wide}
\eeq
Then it becomes identical to that found in the PS/ES models, with the density contrast in top-hat spherical collapse $\delta\F$ replaced by that in Gaussian ellipsoidal collapse $\delta$ and the background density $\delta\F\m$ replaced by $q\delta\m$. 

Truncating the series (\ref{exp}) at first order, we have 
\beqa
\delta\pk(R,\delta|\delta\bb)\approx B_{1} (R,\delta) \delta\bb~~~~~~~~~~~~~~~~~~~~~~~~\label{rel}\\
B_{1} (R,\delta)= -q \frac{\partial \ln n\pk(R,\delta)}{\partial \delta}.~~~~~~~~~~~~~~~~~~~~~
\label{Lag}
\eeqa
Since the PBS approach guarantees that, for $R'>3R$, the autocorrelation of the relation (\ref{rel}) yields $\xi\pk(r)\approx (B_1)^2\xi\m(r)$, where $\xi\pk(r)$ and $\xi\m(r)$ are the peak and matter correlation functions on scale $R$ and $R'$, respectively, and, for $r>4R'$, $\xi\m(r)$ is also nearly the same as the matter correlation function on scale $R$, we have that the local linear bias $B_{1}$ given by equation (\ref{Lag}) is at the same time the perturbative linear bias. In Appendix \ref{B} we give its detailed analytic expression. At intermediate and high masses ($\log \nu\ga 0.1$), it takes the simple form
\beq
B_1(R,\delta)= q\frac{\nu-\gamma\,\lav x^2 \rav/\lav x \rav}{\sigma_0\,(1-\gamma^2)},
\label{simple1}
\eeq
which tends to $B_1= q\nu/\sigma_0$ at very high masses where $\lav x^2\rav$ approaches $\lav x\rav^2$ with $\lav x\rav\approx \gamma\nu$. See the similarity with $B_1\F=\nu\F/\sigma_0\F$ in the PS and ES models, respectively (the latter also for very large masses).

It is worthwhile noting that, for $M\vir$ masses, $B_1$ expressed as a function of $\nu\F$ (using eq.~[\ref{nus}]) with fixed $\delta\F$ instead of $M$ is nearly universal, i.e. nearly a function of $\nu\F$ alone. Indeed, since for $M\vir$ masses $R$ does not depend on $t$ (see Sec.~\ref{CUSP}), $\partial \ln n\pk/\partial\delta=\partial \ln f/\partial \delta$, where $f\equiv M\, n\pk \der R/\der \ln \sigma_0^{-1}$ is the mass fraction in peaks per infinitesimal $\ln \sigma_0^{-1}$ or multiplicity function. As shown in \citet{Jea14b}, for $M\vir$ masses, $f$ expressed as a function of $\nu\F$ is nearly a function of $\nu\F$ alone ($\sigma_2R/\sigma_0$ and $M/R_*^3$ are roughly constant) and, since $r_\delta$ is nearly the same in all cosmologies, $\partial \ln n\pk/\partial \delta\approx \partial \ln [f \partial \ln\nu/\partial \ln\nu\F]/\partial \nu\F \times 1/[r_\delta\sigma_0\F]$ is also essentially a function of $\nu\F$ alone (for fixed $\delta\cc\F$). This theoretical inference is supported by simulations.\footnote{For instance, T+10 found that the bias is nearly universal for all mass definitions they tried, but the best result was found for $M\vir$ masses (see their Table 3).} And, given the recursive relation (\ref{Bi}), the same conclusion holds for all $B_i$ parameters.

The Eulerian linear halo bias $b_1$ can be readily obtained from its Lagrangian counterpart $B_1$. Since $n\pk$ is corrected for nesting, the peak overdensity $\delta\pk(R,\delta)$ at $\ti$ (eq.~[\ref{rel}]) coincides with the halo overdensity $\delta\h(M,t)$ at $t$, that is we must not worry about all minor and major mergers taking place between $\ti$ and $t$. Therefore, in the PBS approximation, the Eulerian overdensity of halos with mass $M$ at $t$ is simply the invariant (Lagrangian) overdensity of peaks lying on the background plus the density contrast of the background itself at $t$,
\beq
\delta\h(M,t)=\delta\pk[R(M,t),\delta(t)|\delta\m(t)]+\delta_{\rm m}(t).
\label{pre}
\eeq
Taking into account the relation (\ref{rel}) and dividing equation (\ref{pre}) by $\delta\m(t)$, we are led to the Eulerian (local or perturbative\footnote{The autocorrelation of $\delta\h(M,t)$ leads to $\xi\h(r)=b_1^2(M,t)\xi\m(r)$, where $\xi\h(r)$ and $\xi\m(r)$ are, at large $r$, very nearly equal to the halo and matter correlations at $t$ smoothed on scale $M$.}) linear bias of halos with $M$ at $t$
\beq
b_{1}(M,t)\equiv \frac{\delta\h(M,t)}{\delta_{\rm m}(t)}=1+B_{1} (M,t).
\label{bias}
\eeq 

The higher order Eulerian bias parameters are harder to derive. The relation (\ref{bias}) for the linear bias is, as mentioned, independent of the evolution of $\delta\m$, but this is not the case at higher orders. In the general case, $\delta\pk[R(M(t),\delta(t)|\delta\m(t)]$ includes terms with powers of $\delta\m$ (eq.~[\ref{exp}]), so by dividing the relation (\ref{pre}) by $\delta\m$, the background density does not simplify and the resulting $b_i$ depend on the evolved background density $\delta\m(t)$. When it is calculated in gravitational Perturbation Theory using top-hat spherical collapse, one is led to the relations \citep{Wea15}
\beqa
b_1\F(M,t)=1+B_1\F(M,t)~~~~~~~~~~~~~~~~~~~~~~~~~~~~~~~~~~~~~~\nonumber\\
b_2\F(M,t)=\frac{8}{21}B_1\F(M,t)+B_2\F(M,t)~~~~~~~~~~~~~~~~~~~~~~~\nonumber\\
b_3\F(M,t)\!=\!-\frac{796}{1323}B_1\F(M,t)\!-\!\frac{13}{7}B_2\F(M,t)\!+\!B_3\F(M,t)\!\!\!~~\nonumber\\
...~~~~~~~~~~~~~~~~~~~~~~~~~~~~~~~~~~~~~~~~~~~~~~~~~~~~~~~~~~~~~~~~~~~~
\label{localE}
\eeqa
allowing one to calculate the higher order Eulerian local parameters from the Lagrangian ones in top-hat smoothing. But the predictions of CUSP have been derived using Gaussian smoothing, so we cannot use these relations to obtain the higher order Eulerian local bias parameters. We should first convert the predicted Gaussian-smoothed Lagrangian bias parameters $B_i$ to the top-hat ones, $B_i\F$, which is not obvious at all. 

The overdensity $\delta^f$ at a point {\bf x}\ in the Gaussian random density field at $t$ smoothed with filter $f$ of scale $M$ (or $R$) normalized to the corresponding rms density contrast, $\sigma_0^f$, cannot depend on the scale because its autocorrelation is always unity. As a consequence, the conversion of $\delta^f/\sigma_0^f$ from Gaussian to top-hat smoothing cannot depend on the scale either. Consequently, we have $(\delta\h/\sigma_{\rm 0h})/(\delta\m/\sigma_0)=(\delta\h\F/\sigma\F_{\rm 0h})/(\delta\m\F/\sigma_{\rm 0}\F)$ or, equivalently, 
\beq
\frac{\delta\h\F}{\delta\m\F}=\frac{\delta\h}{\delta\m}\frac{\sigma_0\sigma_{\rm 0h}\F}{\sigma_0\F\sigma_{\rm 0h}},
\label{conv}
\eeq
where $\sigma_{\rm 0h}$ and $\sigma_{\rm 0h}\F$ stand for the rms peak number overdensity in Gaussian and top-hat smoothing, respectively. Taking the Eulerian local expansion for Gaussian smoothing (eq.~[\ref{exp}]) divided by $\delta\m$,
\beq
\frac{\delta\h(M,t)}{\delta\m(M,t)}\approx 1+B_1(M,t) +\sum_{i\ge 2} b_i(M,t) \delta_m^{i-1},
\eeq
multiplying it by the conversion factor (\ref{conv}) leading to $\delta\h\F/\delta\m\F$, identifying each order term with that in the similar relation in top-hat smoothing, we are led at first order to $(\sigma_{\rm 0h}\sigma_0\F)/(\sigma_0\sigma_{\rm 0h}\F)\approx 1$ and, hence, $B_1\F=B_1$, implying $b_1=b_1\F$. Unfortunately, even though the highest order Lagrangian bias parameters entering the $b_i$ and $b_i\F$ terms are $B_i$ and $B_i\F$, respectively, which leads to $B_i\F(M,t)=B_i(M,t)(\delta\bb/\delta\bb\F)^{i-1}$, the unknown ratio $\delta\bb/\delta\bb\F$ (possibly dependent on $M$) does not allow one to determine $B_i\F$ for $i>1$. Therefore, the only Eulerian (or Lagrangian) bias parameter predicted by CUSP using Gaussian smoothing that can be checked against simulations using top-hat smoothing is the linear one. 

\section{Comparison with Simulations}\label{sim}

For this comparison we will use the ``empirical'' Eulerian linear biases obtained by means of two {\it independent} techniques: the correlation of simulated halos and `separate-universe' simulations, by T+10 and LWBS, respectively. The results of both techniques coincide within the statistical error, which gives strong confidence on them. Nevertheless, they could still be affected by the same (or very similar) systematics due to the common halo-finding algorithm they use. Thus, before proceeding to the comparison, it is worth reminding the limitations of these kinds of algorithms.

The FOF algorithm is known to cause a substantial amount of spurious halo splitting or grouping, depending on the linking length $l$ used. In principle, examining the results arising from different values of $l$, it should be possible to assess the impact of these effects. Unfortunately, all studies of halo bias using this algorithm adopt $l=0.2$. The SO algorithm with overdensity $\Delta$ is sensitive to the roughly spherical shape of halos, which tends to diminish those spurious effects, and uses a specific procedure that minimizes splitting (though likely busts grouping).\footnote{When one peak is found inside the radius of another, the SO finding algorithm keeps the latter and rejects the former as a separate object.} Consequently, the SO algorithm is clearly preferable. Nevertheless, it does not fully avoid those spurious effects, which may alter the empirical bias relative to the ideal (unaffected) one predicted by CUSP. In this sense, the study by T+10 of the effect on the bias of different values of $\Delta$ will be of much help.

\begin{figure}
\includegraphics[scale=1.3, bb= 45 10 400 198]{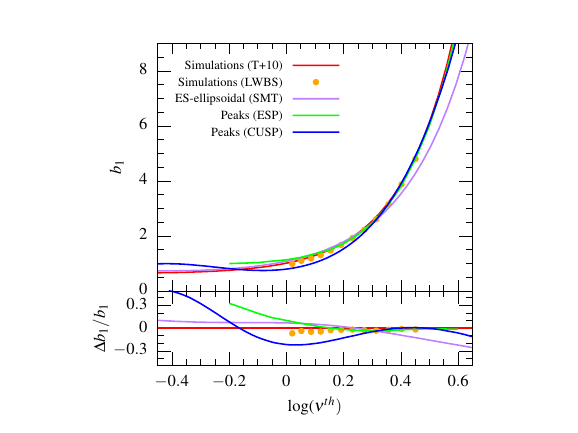}
 \caption{Eulerian linear halo bias predicted by CUSP for $M\vir$ masses at $z=0$ in the {\it WMAP7} cosmology, compared to the results of simulations and other theoretical predictions. The relative errors with respect to the T+10 bias are shown in the lower panel.}
\label{f5}
\end{figure}

The large-scale linear biases provided by T+10 and LWBS were obtained using the SO algorithm with $\Delta=\Delta_{200}$. However, we prefer to deal with the bias corresponding to $\Delta=\Delta\vir$ corresponding to $M\vir$ masses for the following reasons: 1) as shown by \citep{Jea14b}, for roughly spherical objects as halos, the SO algorithm with $\Delta\vir$ is equivalent to the FoF one with $l=0.2$, so our results will also be comparable (except for the different amount of spurious halo splitting and grouping) to the empirical bias obtained from simulations using the FoF halo-finding algorithm, 2) the scale $R$ of peaks at $\ti$ corresponding to halos with $M\vir$ masses is a function of $M$ alone (it does not depend on $t$; see Sec.~\ref{intro}), which simplifies the calculations in Appendix \ref{D} and 3) as mentioned, the bias parameter for $M\vir$ masses expressed as a function of $\nu\F$ is nearly universal, so we must not worry about small differences in the cosmologies and redshifts used in simulations. Therefore, we have converted the results provided by T+10 and LWBS from $\Delta=200$ to $\Delta\vir$ by means of the relation (6) provided by T+10. (In the case of LWBS results the conversion has been achieved by keeping the same fractional deviation from T+10 data in $\Delta\vir$ as in $\Delta_{200}$.) 

The linear bias provided by T+10 is in the form of an analytic fit (with a quoted accuracy of $\sim 5$\%) to the stacked data in the mass range $-0.1\la \log(\nu\F)\la 0.55$ of several simulations (encompassing different volumes) carried out in slightly different $WMAP$ cosmologies and redshifts (from $z=0$ to $z=2.5$), using the same fixed value of $\delta\F\cc=1.686$. This stacking is justified by the above mentioned approximate universality of the bias in these conditions. The liner bias provided by LWBS is in the form of discrete points corresponding to a binning in halo masses, so, strictly speaking, it is rather a convolution of the real bias with the window function of the mass bin (LWBS). However, given the small bin size, this makes a small difference we neglect here. This bias corresponds to $z=0$ in a cosmology of the {\it WMAP} kind. Therefore, the theoretical linear bias predicted by CUSP has been calculated at $z=0$ in the {\it WMAP7} cosmology and expressed as a function of $\nu\F$ (by means of the relation [\ref{nus}]) like the empirical biases it is to be compared to. 

The comparison is shown in Figure \ref{f5}. Contrarily to the bias predicted in the ES model, those predicted by CUSP and ESP match the empirical bias at high masses. We remind that both formalisms are built in the peak model framework and, since at high masses there is no peak nesting, they only differ in the mass and ellipsoidal collapse time of Gaussian peaks they adopt, apart from the different filter they use, which is irrelevant for $b_1$. At intermediate masses, however, the CUSP bias decreases with decreasing mass more steeply than all the remaining biases, including the ESP one, until it reaches a minimum at $\log \nu\F \sim -0.1$ and increases again. As a consequence, it slightly undulates around the other monotonically decreasing curves, which are underestimated by about $25$\% at intermediate masses and overestimated by about $50$\% at $\log\nu\F\sim -0.40$. The existence of that minimum (found regardless of the value of $q$), if real, would imply that the usual behavior of more massive halos being more strongly clustered is reversed at low masses, which is quite surprising. Whether this reflects the limits of the peak model at low masses due to the shear of the mass distribution around peaks is unclear. However, the collapse of protohalos anchored at the center of peaks is less affected by shear than the collapse of protohaloes at fixed points as in the PS/ES models, so the halo-peak correspondence should hold down to very low masses, as supported by the well predicted halo MF there.

In any event, the significance of this minimum cannot be assessed on the basis of the departure from the empirical bias obtained by T+10 at those low masses as it is just an extrapolation of the fit to the real data (in the same range as the LWBS points). We can only try to estimate the error of our prediction. The analytic expression of $n\pk$ matches its numerical counterpart at $z \le 2$ to better than 0.02\% (Fig.~\ref{f4}) and $\delta(z)$ is also accurate to better than 1\% (App.~\ref{C}), so the analytic derivative $\partial\ln n\pk/\partial \delta$ should also be accurate to about 3\%. But, as mentioned in Section \ref{unconditional}, the numerical peak number density solution of equation (\ref{nnp}) is only accurate at the 5\% level, so the analytic derivative $\partial\ln n\pk/\partial \delta$ fitting it could have, after all, a substantial error. According to the relation 
\beq
-\frac{\partial \ln n\pk}{\partial \delta}=\frac{\partial \ln [M \partial R/\partial \nu]}{\partial \delta}-\frac{\partial \ln f}{\partial \delta},
\eeq
the relative error of $b_1$ at $\log \nu\F <0$, where $b_1$ is about unity, is essentially equal to the absolute error of $\partial \ln f/\partial \delta$. Since the multiplicity function $f$ is nearly universal \citep{Jea14b}, the analytic derivative $\partial \ln f/\partial \delta$ is close to zero like its empirical counterpart \citep{Wea06}.\footnote{These authors used the FoF halo finding algorithm with $l=0.2$, which, as mentioned, coincides for spherical halos with the SO algorithm with $\Delta\vir$, so their MF corresponds to $M\vir$ masses.} Consequently, an upper bound of the difference between both measuring the error of the analytic value is given by the absolute value of the numerical derivative $\partial \ln f/\partial \delta$, equal to $\sim 0.02$ (see Fig.~1 of \citealt{Jea14b}). Therefore, the estimated relative error of the theoretical bias at those masses is less than $\sim 2$\%. This makes an upper bound of the total error of about 5\%, which is clearly insufficient to make the minimum be insignificant. 

At $\log \nu\F >0$ where there is essentially no peak nesting, the theoretical bias given by equation (\ref{Lag}) with $n\pk=N\pk$ is much more accurate, so the difference between the theoretical and empirical biases at that the mass range covered by simulations is certainly significant. Does this mean that the prediction of CUSP is deficient there, or is the empirical bias affected by some spurious effect? The analysis by T+10 of the behavior of $b_1$ as a function of $\Delta$ found in simulations suggests that the latter is the right answer. Indeed, when $\Delta$ diminishes, halos become more massive, while they are identically clustered, so $b_1$ should shift horizontally towards small masses and, given its slope, vertically upwards at fixed mass. This expected behavior is satisfied by the bias predicted by CUSP (dependent on the halo mass definition; see Section \ref{CUSP}), but, as seen by application of the relation [6] of T+10, not by the empirical bias which shows the opposite trend at large $\Delta$. That anomalous behavior of the empirical bias is precisely the one we would expect from the effects of halo splitting and grouping: the larger $\Delta$, the smaller halo radii, so the larger the fraction of massive objects that are split in separate subhalos and the smaller the fraction of close halos that are grouped in single objects as well. What is even more compelling, when $\Delta$ increases, the empirical linear bias is found to increase at low masses where $b_1$ is essentially flat and the above mentioned natural horizontal shift cannot lift it upwards. As explained by T+10, this increase is due to the fact that, when $\Delta$ increases, more massive halos are split into their subhalos, which gives rise to spurious intermediate- and low-mass halos that are more clustered than the real halos of that mass. 

Therefore, it is worth checking whether the spurious halo splitting and grouping affecting simulations\footnote{Even though both effects go in the opposite direction, they do not balance each other because splitting acts preferentially at high masses, while grouping does at low masses.} could explain the observed discrepancy. 

\section{Halo Splitting and Grouping}\label{relax}

To that aim we will model those spurious effects and repeat the comparison either with the empirical linear bias corrected for it or with the theoretical one including it. Both procedures are equivalent, but the latter is easier to implement, so we will proceed that way. It is important to remark that, since the exact strength and mass dependence of these effects depend in a convoluted way on the characteristics of each simulation, the model we will build is not intended to be very realistic, but to catch the physics of the problem in enough detail to see whether those effects may cause, indeed, the discrepancy between our predictions and the results of simulations.

Let us start with halo splitting. Halos having just undergone a major merger so that the subhalos arising from the halo progenitors are still very apparent are the most susceptible to be split by the SO algorithm in separate objects. Consequently, in order to include halo splitting in the peak bias we must monitor the peak process tracing halo mergers. This is possible thanks to CUSP \citet{SM19}. Specifically, we will use the results derived in \citet{MSS96}, but focusing on {\it non-nested} peaks, which simplifies the treatment.

The fraction of non-nested peaks with $\delta$ at $R$ appearing per infinitesimal density contrast around $\delta_{\rm a}$ as a consequence of halo mergers is the solution of the differential equation
\beq
\frac{\partial f^{\rm a}(R,\delta,\delta_{\rm a})}{\partial \delta_{\rm a}}=r^{\rm a}(R,\delta_{\rm a}),
\label{ff}
\eeq
where 
\beq
r^{a}(R,\delta)\!=\! \frac{\partial \ln n\pk}{\partial \delta}\!-\! \frac{\Sigma(R,\delta)}{R}\!\left[\frac{\partial \ln n\pk}{\partial  \ln R}\!+\!\frac{\partial \ln\Sigma}{\partial  \ln R}\!\right]\!
\label{aa}
\eeq
is the average appearance rate per infinitesimal density contrast $\delta_{\rm a}$ of {\it non-nested} peaks with $\delta$ at $R$, being $\Sigma(R,\delta)=\sigma_0/[\sigma_2R\lav x\rav]$ (see eq.~[C8] in \citealt{MSS96}). Note that equation (\ref{aa}) differs form the expression (17) in \citet{MSS96} holding for all peaks in that there is no term correcting for peaks becoming nested.  

Given that, by continuity, $f^{\rm a}(R,\delta,\delta_{\rm a})$ vanishes at $\delta_{\rm a}=\delta$, equation (\ref{ff}) can be readily integrated for variable $\delta_{\rm a}$ between $\delta$ and $\delta_{\rm a}$ slightly larger than $\delta$, leading to
\beq
f^{\rm a}(R,\delta,\delta_{\rm a})
\approx (\delta_{\rm a}-\delta)\,r^{\rm a}(R,\delta).
\label{fa}
\eeq
Therefore, the fraction $\Delta n\pk/n\pk$ of halos with $M$ at $t$ formed in major mergers during the last $k_{\rm s}$ crossing times of the system, $\tau\cc(t)=[\pi G\bar\rho(t)]^{-1/2}$, with $k_{\rm s}$ small enough for those objects to be split by the SO algorithm in their progenitors, is the fraction of peaks with $\delta$ at $R$ appearing in the corresponding $\delta$-interval, i.e. between $\delta+{\Delta_{\rm s}}(\delta,k_{\rm s})\equiv \delta[t-k_{\rm s}\tau\cc(t)]$ and $\delta=\delta(t)$, 
\beqa
\frac{\Delta n\pk(R,\delta)}{n\pk(R,\delta)}\!=\!\int_\delta^{\delta+{\Delta_{\rm s}}(\delta,k_{\rm s})}\!\!\!\!\! \der \delta' f^{\rm a}(R,\delta,\delta')~~~~~~~~\nonumber\\
\approx\frac{1}{2}\Delta_{\rm s}^2(\delta,k_{\rm g})r^{\rm a}(R,\delta).~~~~~~~~~~~~~
\label{D0}
\eeqa
Clearly, $k_{\rm s}$ depends on $\Delta$, but if halos are defined by means of the SO finding algorithm with $\Delta=\Delta\vir(t)$, $k_{\rm s}$ does not depend on $t$. Indeed, in this case the overdensity used to identify halos varies with $t$ at the same rate as the overdensity of virialized halos, so the appearance relative to such a gauge of a halo after any given number of crossings of its progenitors is the same at whatever cosmic time. 

Lastly, taking into account that halos with $M$ at $t$ caught at less than $k$ crossing-times after the last merger are split into their main subhalos tracing the progenitor halos and that major mergers are binary with the mass of the two progenitors equal to about half the mass of the final object (e.g. \citealt{Rea01}), equation (\ref{D0}) leads to the following effective (i.e. including halo splitting) number density of peaks with $\delta$ at $R$ lying on a background with $\delta\m$ 
\beqa
n\pk^{\rm s}(R,\delta-q \delta\m)= n\pk(R,\delta-q\delta\m)\,C_{\rm s}(R,\delta)
\label{new3}\,~~~\nonumber\\
C_{\rm s}(R,\delta)\approx 1-\frac{\Delta n\pk(R,\delta)}{n\pk(R,\delta)}+2\frac{\Delta n\pk(R_{\rm s},\delta)}{n\pk(R,\delta)},
~~\label{aver}
\eeqa
where $R_{\rm s}\equiv 2.7^{1/3}R$. This value of $R_{\rm s}$ presumes that subhalos into which a halo is split are about $25$\% less massive than the real halo progenitor due to the fact that the latter are tidally stripped during the merger, so the mass of the halo being split is $\sim 2.7$ instead of two times the mass of each spurious halo added from the splitting. (See App.~\ref{D} for the detailed analytic expression of $C_{\rm s}$.) 

But expression (\ref{aver}) is not enough: following the same derivation as in Section \ref{LB} with $n\pk(R,\delta-q\delta\m)$ replaced by $n\pk^{\rm s}(R,\delta-q\delta\m)$ leaves the Lagrangian linear bias unchanged. This is because the peak appearance rate used in equation (\ref{aver}) is an {\it average over all background densities}, while halo mergers are more frequent in high density regions than in low ones. This means that, instead of using the average peak appearance rate, we must use the appearance rate of peaks conditioned to lie in a background $\delta\m$, $r^{a}(R,\delta|\delta\m)$, defined just as $r^{a}(R,\delta)$ (eq.~[\ref{aa}]) but with the conditional peak number density, $n\pk(R,\delta-q\delta\m)$, instead of the unconditional one. Then we obtain
\beq
n\pk^{\rm s}(R,\delta-q\delta\m)=n\pk(R,\delta-q\delta\m)\, C_{\rm s}(R,\delta,\delta\!-\!q\delta\m),
\label{nk}
\eeq
with $C_{\rm s}(R,\delta,\delta\!-\!q\delta\m)$ defined as $C_{\rm s}$ in equation (\ref{aver}), but with $\delta$ replaced by $\delta-q\delta\m$, except in the factor $\Delta_{\rm s}^2(\delta,k_{\rm s})$, which remains unchanged. 

Once the effective conditional peak number density has been fixed, the same development as in Section \ref{LB} leads to the desired relation between the bias parameter including halo splitting, $B_1^{\rm s}(R,\delta)$, and the one derived above, $B_1(R,\delta)$, not including it, 
\beq
B_1^{\rm s}(R,\delta)
\approx B_1(R,\delta)-q\frac{\partial \ln C_{\rm s}}{\partial \delta},
\label{noaver}
\eeq
where the $\delta$-derivative of $\ln C_{\rm s}$ affects the last argument of this function only. 

Let us now turn to the halo-grouping effect. Grouping is expected to mostly affect close halos that are going to merge in the next $k_{\rm g}$ times $\tau\cc(t)$. $k_{\rm g}$ is also expected to be of order of unity and different in general from $k_{\rm s}$ because, as we will see, splitting and grouping do not balance each other. Once again, major mergers are binary with the mass of progenitors about half the mass of the final halo, so the situation is symmetric to that found in halo splitting, with the difference that $k_{\rm g}\tau\cc(t)$ is now added and not subtracted to $t$ and that the number density of halos with mass $M$ increases at the expense of halos with mass about half instead of twice $M$. Consequently, the fraction $\Delta n\pk/n\pk$ of peaks with $R$ that will disappear from $\delta$ to $\delta-\Delta_{\rm g}(\delta,k_{\rm g})$ due to the merger of halos with $M$ in the interval between $t$ and $t+k_{\rm g}\tau\cc(t)$, respectively, is essentially twice the fraction of peaks appearing from $\delta$ to $\delta-\Delta_{\rm g}(\delta,k_{\rm g})$ with $R'_{\rm s}=2^{1/3}R$ (the tidal stripping of the merging objects must not be taken into account now), 
\beq
\frac{\Delta n\pk(R,\delta)}{n\pk(R,\delta)}\approx \Delta_{\rm g}^2(\delta,k_{\rm g})r^{\rm a}(R'_{\rm s},\delta).
\label{D0bis}
\eeq
We then arrive at the following effective (i.e. including halo-grouping) number density of peaks with $\delta$ at $R$ subject to lying on a background with $\delta\m$ 
\beqa
n\pk^{\rm g}(R,\delta-q\delta\m)= n\pk(R,\delta-q\delta\m)\,C_{\rm g}(R,\delta)
\label{new3tris}\,~~~~~~~~~~\nonumber\\
C_{\rm g}(R,\delta,\delta\!-\!q\delta\m)\!\approx\! 1\!-\!\frac{\Delta n\pk(R,\delta)}{n\pk(R,\delta)}\!+\!\frac{1}{2}\frac{\Delta n\pk(R_{\rm g},\delta)}{n\pk(R,\delta)}.~~~~
\label{noavertris}
\eeqa
where $R_{\rm g}\equiv R/2^{1/3}$. (See App.~\ref{D} for the detailed analytic expression of $C_{\rm g}$.) Strictly speaking, near the halo mass limit of the simulation, grouping only subtracts halos; no halo is added from the grouping of less massive objects. Consequently, the second term on the right in equation (\ref{noavertris}) must be taken null, though this makes a negligible difference in the results. 

\begin{figure}
\includegraphics[scale=1.3, bb= 45 10 400 198]{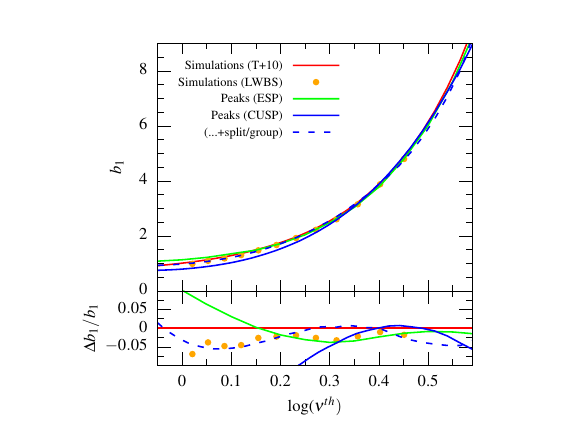}
 \caption{Same as Figure \ref{f5} but with an additional version of the Eulerian linear halo bias predicted by CUSP including the modelled effects of spurious halo splitting and grouping (blue dashed line).}
\label{f6}
\end{figure}

Including halo grouping {\it after} halo splitting in the number density of peaks lying on the background $\delta\m$ leads to the relation
\beqa
n\pk^{\rm sg}(R,\delta\!-\!q \delta\m)\!~~~~~~~~~~~~~~~~~~~~~~~~~~~~~~~~~~~~~~~~~~~\nonumber\\
= n\pk^{\rm s}(R,\delta\!-\!q\delta\m)\frac{n\pk(R,\delta\!-\!q\delta\m)}{n\pk^{\rm s}(R,\delta\!-\!q\delta\m)}C_{\rm g}(R,\delta,\delta\!-\!q\delta\m),~~~~
\label{noaver4}
\eeqa
which in turn implies a Lagrangian linear bias of the form
\beq
B_1^{\rm sg}(R,\delta)
\!\approx\! B_1(R,\delta)-q\left[\frac{\partial \ln C_{\rm s}}{\partial \delta}+\!\frac{\partial \ln C_{\rm g}}{\partial \delta}\right]\!.
\label{new4}
\eeq
Note that equation (\ref{new4}) is symmetric with respect to splitting and grouping, implying that, as expected, the order of the inclusion of these effects does not matter. (See App.~\ref{D} for the detailed analytic expression of $\partial \ln C_{\rm s}/\partial \delta$ and $\partial \ln C_{\rm g}/\partial \delta$.) 

In Figure \ref{f6} we show the effect of including the modelled spurious halo splitting and grouping in the linear bias predicted by CUSP. The solution plotted is for the values of $k_{\rm s}$ and $k_{\rm g}$ equal to $0.95\pm 0.05$ and $1.45\pm 0.05$, respectively, giving the best fit (i.e. the minimum $\chi$-squared value) to the LWBS data. Note that these values are close to unity as expected, though, for the reasons mentioned at the beginning of this Section, they may somewhat deviate, of course, from the values that would be found in the implementation of the same strategy in any specific simulation. The comparison is restricted to the mass range $-0.05<\log(\nu\F)<0.55$ covered by the real data. We see that the theoretical bias reproduces now the T+10 fit to better than $\sim 5$\% and the LWBS points to better than $\sim 3$\%. Furthermore, the small undulation of the theoretical bias around the T+10 fit that remains in that mass range is similar to that shown by the data in the individual simulations carried out by these authors (see the different colored points in Fig.~1 of T+10). From these results we conclude that: i) even a low level of halo splitting and grouping as corresponding to such small values of $k_{\rm s}$ and $k_{\rm g}$ has a notable impact in $b_1$, and ii) taking into account these effects, the predicted bias is in excellent agreement with the results of simulations.

\section{SUMMARY AND CONCLUSIONS}\label{dis}

DM halos are complex biased tracers of matter: they show a primary bias, i.e. the more massive halos, the more clustered, and a secondary (or assembly) bias, i.e. halos with the same mass but different internal properties are also differently clustered. While the origin of the secondary bias is unclear, the primary bias is well-understood, though not satisfactorily reproduced by current galaxy formation models. Specifically, the halo bias predicted in the PS or ES models deviates from the results of simulations at high masses, while the opposite is true for that predicted by ESP in the peak model. 

In the present Paper we have used the CUSP formalism to carry out an alternative derivation of the halo bias in the peak model. Following DCSS, we have calculated in the PBS approximation the Lagrangian local peak bias parameters by differentiation of the conditional number density per infinitesimal scale of peaks lying on a background, properly correcting it for peak nesting. 

The resulting analytic expressions are simple and similar to those obtained in the PS/ES models, but depend on the halo mass definition adopted, as expected from simple logical arguments and similarly to the results of simulations. The latter show, however, some anomalies due to spurious halo splitting and grouping caused by the SO halo-finding algorithm. 

While current simulations (and all previous models) use top-hat smoothing, CUSP uses Gaussian smoothing. Nevertheless, the linear bias is the same in all filters, so its comparison with the results of simulations is still possible. This way we find that the predicted Eulerian linear halo bias agree with the bias of simulated halos at high masses though a small departure is observed at lower masses. However, we have shown that this departure is likely due to the above mentioned spurious effects in simulations. 

Thus, the main conclusions of this work are: i) the spurious halo splitting and grouping caused by the usual halo-finding algorithms can significantly alter (relative to the statistical error) the bias found in simulations, and ii) CUSP predicts a linear bias in excellent agreement with the results of simulations within the uncertainty due to the previous effect. 

The latter conclusion is particularly remarkable compared to the achievements of ESP using the same conditional peak number density though uncorrected for peak nesting. While the mass and time of ellipsoidal collapse of Gaussian peaks in ESP, chosen in a physically motivated way, harbor a few parameters that are fixed in a simulation-guided manner, in CUSP they are fully determined from first principles and with no free parameters.

In a forthcoming Paper we show that CUSP also explains the secondary (or assembly) bias.

\begin{acknowledgments}
This work was funded by the Spanish MCIN/AEI/ 10.13039/501100011033 through grants CEX2019-000918-M (Unidad de Excelencia `Mar\'ia de Maeztu', ICCUB) and PID2022-140871NB-C22 (co-funded by FEDER funds) and by the Catalan DEC through the grant 2021SGR00679.
\end{acknowledgments}

{}

\begin{appendix}

\section{Refining the density contrast for ellipsoidal collapse}\label{C}

The density contrast for ellipsoidal collapse at $t$, $\delta(t)$, is equal to the homologous quantity in spherical collapse, $\delta\F(t)$, times the factor $r_\delta(t)$. In previous works, we took $r_\delta(t)$ given by the simple expression  
\beq
r_\delta(t)=\frac{a^d(t)}{D(t)}.
\label{old}
\eeq
This expression was obtained in \citet{Jea14a} by fitting the ratio $\delta(t)/\delta\F(t)$ found at redshifts $z=0$, 5, 10 and 20 covering the whole relevant range of baryon-trapping halos. It is sufficient, indeed, both at $z\ga 5$, where $r_\delta(t)$ is very sensitive to the value of (cosmology-dependent) coefficient $d$ as well as at $z=0$, where it is always equal to $=1/D_0$, where $D_0\equiv D(t_0)$, regardless of the exact value of $d$. However, near $z=0$, there is a blind spot where $d$ could significantly deviate from the value found at $z\ga 5$. In all previous works, we used CUSP to infer the properties of halos at $z=0$, so expression (\ref{old}) was enough. However, in the present Paper, we need to be a more precise as we deal with $r_\delta(t)$ not only at $z=0$ but also near to it. 

It is thus convenient to adopt the following refinement of expression (\ref{old})
\beq
r_\delta(t)=\frac{a^{d{\cal D}(t)}(t)}{D(t)},~~~~~~~~~~~~~~~~~~~~~~~~~~~~~~~ {\rm where}~~~~~~~~~~~~~~~~~~~~ ~~~~~~~~~~  {\cal D}(t)=1-d_0a^{0.435a^{-1}(t)}(t),
\label{renew}
\eeq
which fits not only the values of $\delta(t)/\delta\F(t)$ at $z=0$ and $z\ga 5$, but also in the blind spot (specifically, at $z=0.5$ and $z=2$) to better than 1\%.

\section{Analytic Expression of the Peak-Nesting Correction}\label{A}

Multiplying and dividing by $N\pk(R,\bar\delta)$, where $\bar\delta=\bar q\delta$, with $\bar q=1-q$, inside the integral on the right of equation (\ref{nnp}) and approximating the numerator by its zero-order Taylor expansion at a suited background scale $R\e$ larger than $R$, equation (\ref{nnp}) becomes 
\beq
n\pk(R,\delta)\approx N\pk(R,\delta)
-N\pk(R,\bar q\e\delta)\int_{R}^{\infty}\der R' \frac{N\pk(R,\delta|R',\delta)}{N\pk[R,[\bar q(R')\delta]}f(R',\delta),
\label{B1bis}
\eeq
where $\bar q\e\equiv \bar q(R\e)$, $f(R,\delta)\equiv M(R,\delta)/\bar\rho(\ti) n\pk(R,\delta)$ is the mass fraction in non-nested peaks with $\delta$ at $R$ per infinitesimal $R$ corrected for nesting. Thanks to the fact that at $R'>3R$ $N\pk(R,\delta|R',\delta)$ becomes $N\pk[R,\bar q(R')\delta]$ (while at $R'=R$ it vanishes), the ratio $N\pk(R,\delta|R',\delta)/N\pk[R,\bar q(R')]\delta]$ appears to be very nearly equal, for all $R$ and $\delta$, to $J(R'/R)\equiv 1-\exp[1-(R'/R)^2]$, with $J(1)=0$. Thus, defining the nesting correction factor $C_{\rm n}\equiv n\pk/N\pk$ so that $f(R,\delta)= f\uu(R,\delta)C_{\rm n}(R,\delta)$, where $f\uu(R,\delta) \equiv M(R,\delta)/\bar\rho(\ti) N\pk(R,\delta)$ is the nesting-uncorrected mass fraction of peaks with $\delta$ at $R$, we are led to
\beq
\frac{f\uu(R,\delta)}{f\up(R,\bar\delta)}\left[C_{\rm n}(R,\delta)-1\right]\approx -\int_{R}^{\infty}\der R'\, J\!\left(\frac{R'}{R}\right) f\uu(R',\delta) C_{\rm n}(R',\delta),
\label{B2}
\eeq
with $f\up(R,\bar\delta)\equiv M(R,\delta)/\bar\rho(\ti) N\pk(R,\bar\delta)$ and $\bar\delta=\bar q\e\delta$, or equivalently
\beq
\frac{\partial}{\partial R}\!\left\{\frac{f\uu(R,\delta)}{f\up(R,\bar\delta)}[C_{\rm n}(R,\delta)-1]\right\}
\approx  2\int_{1}^{\infty}\der \!\left(\frac{R'}{R}\right) \left(\frac{R'}{R}\right)^2 {\rm e}^{1-\left(\frac{R'}{R}\right)^2}f\uu(R',\delta) C_{\rm n}(R',\delta).
\label{B3}
\eeq 
Taking advantage of the exponential in the integrand on the right of equation (\ref{B3}), we can approximate $f\uu(R',\delta) C_{\rm n}(R',\delta)$ by its Taylor expansion around $R$ to first order in $(1-R/R')$, which leads to
\beq
\frac{\partial}{\partial R}\!\left\{\frac{f\uu(R,\delta)}{f\up(R,\bar\delta)}[C_{\rm n}(R,\delta)-1]\right\}
\approx  c_0 f\uu(R,\delta) C_{\rm n}(R,\delta) +c_1 R \frac{\partial (f\uu C_{\rm n})}{\partial R}, 
\label{B4}
\eeq
where $c_0=0.5[2+{\rm e}\sqrt{\pi}\,{\rm erfc}(1)]=1.379$ and $c_1=0.5[2-{\rm e}\sqrt{\pi}\,{\rm erfc}(1)]=0.621$ (but see below). The solution of equation (\ref{B4}) is
\beq
C_{\rm n}(R,\delta)\!\approx \!\frac{H(R,\delta)}{1\!-\!F\up(R,\delta)}\int_0^R \!\!\der R'\,\frac{\partial H^{-1}(R',\delta)}{\partial R'}\,{\rm e}^{\Psi(R,R',\delta)}
{\rm e}^{(c_0-c_1)\int_{R'}^{R}\frac{\der\xi}{f_{\rm p}^{-1}(\xi,\delta)-c_1\xi}}
\label{simple}
\eeq
where
\beq
\Psi(R,R',\delta)\equiv \frac{c_0\!-\!c_1}{c_1}\int_{R'}^R \frac{\der  \xi}{\xi}\,\frac{F\up(\xi,\delta)}{1-F\up(\xi,\delta)}
\label{simple2}
\eeq
$H(R,\delta)\equiv N\pk(R,\bar\delta)/N\pk(R,\delta)$ and $F\up(R,\delta)\equiv c_1 R M(R,\delta)N\pk(R,\bar\delta)/\bar\rho(\ti)$. 

A better approximation can be found, however, by following the same procedure but starting with the Taylor expansion of $N\pk[R,\bar q(R')\delta]$ at a higher order. That would lead to the same equation (\ref{simple})-(\ref{simple2}), but with different values of $c_0$ and $c_1$, dependent on $R\e$ and $\delta$. Of course, that would greatly complicate the calculations, so it is preferable to use expression (\ref{simple}) as a fitting function and adjust the values of $R\e$ and the corresponding parameters $c_0(\delta)$ and $c_1(\delta)$ that would supposedly result at a high enough order. 

Moreover, since $\Psi$ in the function (\ref{simple}) is always smaller than unity, we can make one more step and try the fitting with the simpler expression of $C_{\rm n}$ that results by taking $\exp(\Psi)$ to first order in that quantity. By doing this, we are led to the very practical, accurate enough, analytic expression of $C_{\rm n}$ given by equation (\ref{ap}).

\section{Analytic Expression of the Linear Peak Bias}\label{B}
 
According to the correction factor for nesting (eqs.~[\ref{ap}]-[\ref{ap2}), the $\delta$-derivative of the natural logarithm of the number density of non-nested peaks takes the form
\beqa
\frac{\partial \ln n\pk(R,\delta)}{\partial \delta}
\!=\! \frac{\partial \ln [N\pk(R,\delta)C(R,\delta)}{\partial \delta}\!=\! \frac{\partial \ln N\pk(R,\delta)}{\partial \delta}\!+\!\frac{\partial \ln [1\!-\!H(R,\delta)]}{\partial \delta}\!-\!\frac{\partial \ln [1\!-\!F\up(R,\delta)]}{\partial \delta}\!+\!\frac{\partial \ln[1\!-\!\widetilde C(R,\delta)]}{\partial \delta}~~~~~~~\\
= \frac{1}{1\!-\!H(R,\delta)}J_\delta(R,\delta)-\frac{1\!-\!C(R,\delta)}{1\!-\!H(R,\delta)}\bar J_\delta(R,\delta)-\frac{1\!-\!H(R,\delta)\!-\!C(R,\delta)}{1\!-\!H(R,\delta)}K_\delta(R,\delta)-\frac{\widetilde C(R,\delta)}{1\!-\!\widetilde C(R,\delta)}W_\delta(R,\delta),~~~~~~~~
\label{Lag3}
\eeqa
where 
\beqa
J_\delta(R,\delta)=\frac{\partial \ln N\pk(R,\delta)}{\partial \delta} = \frac{-\nu+\gamma\,\lav x^2 \rav(R,\delta)/\lav x \rav(R,\delta)}{\sigma_0\,(1-\gamma^2)}
\label{J0}~~~~~~~~~~~~~~~~~~~~~~~~~~~~~~~~~~~~~~~~~~~~~~~~~~~~~~~~~~~~~~~~~~~~~\\
\bar J_\delta(R,\delta)=\frac{\partial \ln N\pk(R,\bar \delta)}{\partial \delta} =\bar q\e^2 J_\delta(R,\bar \delta)=\bar q\e^2 \frac{-\bar\nu+\gamma\,\lav x^2 \rav(R,\bar\delta)/\lav x \rav(R,\bar\delta)}{\sigma_0\,(1-\gamma^2)}
~~~~~~~~~~~~~~~~~~~~~~~~~~~~~~~~~~~~~~~~~~~~~~~~~\label{barJ0}\\
K_\delta(R,\delta)=\frac{\partial\ln [c_1(\delta)M(R,\delta)]}{\partial\delta}
=\frac{c_{11}}{c_1(\delta)} -\frac{1}{\delta[1-{\cal S}(\delta)\nu]}\frac{\partial \ln r_\sigma}{\partial\delta}\left(\frac{\der\ln\sigma\F_0}{\der\ln M}\right)^{-1}\!~~~~~~~~~~~~~~~~~~~~~~~~~~~~~~~~~~~~~~~~~~~~\label{u1}\\W_\delta(R,\delta)=\frac{\partial \ln \widetilde C(R,\delta)}{\partial \delta}= I_\delta(\delta)+\frac{1}{1-H(R,\delta)} [\bar J_\delta-J_\delta](R,\delta)+\bigg\lav\!\!\!\bigg\lav \frac{\partial \ln(Cf\uu)}{\partial \delta}\bigg\rav\!\!\!\bigg\rav(R,\delta)\!~~~~~~~~~~~~~~~~~~~~~~~~~~~~~~~~\label{W0}\\
I_\delta(\delta)=  \frac{\der \ln[c_1(\delta)-c_0(\delta)]}{\der \delta}=\frac{c_{11}-c_{01}}{c_1(\delta)-c_0(\delta)}. \,~~~~~~~~~~~~~~~~~~~~~~~~~~~~~~~~~~~~~~~~~~~~~~~~~~~~~~~~~~~~~~~~~~~~~~~~~~~~~~~~~~~~~
\eeqa

In equation (\ref{W0}), double brackets denote average from zero to $R$ for the distribution function $Cf\uu$ duly normalized over that range. This average can be calculated numerically, but we can do better and find an analytic approximation for it. Indeed, defining $\lambda\equiv-\widetilde C/(1-\widetilde C)$ and $\tilde\lambda\equiv -u\widetilde C^\alpha/(1\!-\!\widetilde C)$, with $u=0.30$, $\alpha=0.74$, the term in double brackets satisfies 
\beq
\lambda(R,\delta)\bigg\lav\!\!\!\bigg\lav \frac{\partial \ln(Cf\uu)}{\partial \delta}\bigg\rav\!\!\!\bigg\rav(R,\delta)\approx \tilde \lambda (R,\delta)\frac{\partial \ln(Cf\uu)(R,\delta)}{\partial \delta}+h\lambda (R,\delta),
\label{hu}
\eeq
and $h=-0.36$. These values of $u$, $\alpha$ and $h$ give the best fit to the case of the {\it WMAP7} cosmology, $M\vir$ masses and $z=0$ (where the relation holds to better than 1\% at all masses), but they also provide a good enough fir in the other cases and redshifts $z\la -2.5$ considered in this Paper. Then, since 
\beq
\frac{\partial\ln [Cf\uu](R,\delta)}{\partial\delta}=\frac{1}{1-H(R,\delta)}J_\delta(R,\delta)-\frac{1-C(R,\delta)}{1-H(R,\delta)}\bar J_\delta(R,\delta)+\frac{C(R,\delta)}{1-H(R,\delta)}K_\delta(R,\delta)-\frac{\der \ln c_1(\delta)}{\der\delta}, 
\eeq
equation (\ref{Lag3}) takes the final form
\beq
-\frac{1}{q}B_1(R,\delta)\approx \frac{\partial \ln n\pk(R,\delta)}{\partial \delta}
\approx \iota(R,\delta)J_\delta(R,\delta)+\bar\iota(R,\delta) \bar J_\delta(R,\delta)+\kappa(R,\delta)K_\delta(R,\delta)
+\lambda(R,\delta) L_\delta(R,\delta),
\label{Lag4}
\eeq
where we have introduced the function
\beq
L_\delta(R,\delta)=-\frac{\tilde\lambda(R,\delta)}{\lambda(R,\delta)}\frac{c_{11}}{c_1(\delta)}+\frac{c_{11}-c_{01}}{c_1(\delta)-c_0(\delta)}+h
\eeq
and the coefficients
\beq
\iota(R,\delta)\!\equiv\! \frac{1\!+\!\tilde \lambda(R,\delta)-\lambda(R,\delta)}{1\!-\!H(R,\delta)},~\kappa(R,\delta)\!\equiv\!-1\!+\!C(R,\delta)\frac{1\!+\!\tilde\lambda(R,\delta)}{1\!-\!H(R,\delta)}~~{\rm and}~~\bar \iota(R,\delta)\!\equiv\! 1\!-\!\iota(R,\delta)\!+\!\kappa(R,\delta).\nonumber
\eeq

\begin{figure}
\centering
 \hspace{70pt}
 {\includegraphics[scale=1.0, bb= 100 -5 290 200]{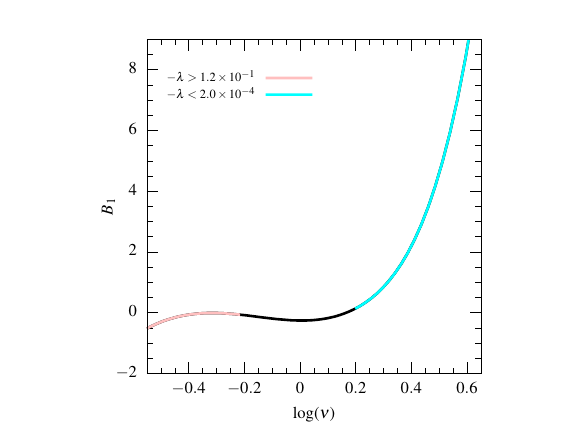}}
 {\includegraphics[scale=1.0, bb= 100 -5 290 200]{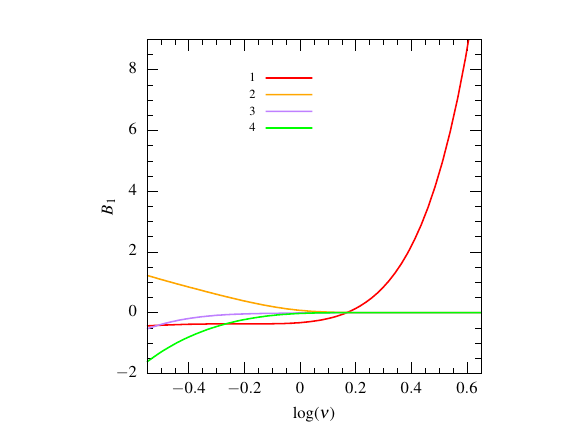}}
 \caption{{\it Left panel}: Lagrangian linear bias predicted by CUSP at $z=0$ in the {\it WMAP7} cosmology and for $M\vir$ masses, indicating the values of $\lambda$ holding at the high-mass and low-mass regions. {\it Right panel}: Contribution of the different terms appearing on the right of equation (\ref{Lag4}), with increasing number from left to right. In particular, curve 1 gives the bias in the absence of peak nesting, i.e. $n\pk=N\pk$.}
 \label{f13}
\end{figure}

In Figure \ref{f13} we plot the resulting Lagrangian bias $B_1$ (eq.~[\ref{Lag4}]), separating the regions of high-mass ($\log \nu > 0.1$) and low-mass halos ($\log \nu < -0.1$) halos and showing the contribution of the different terms on the right of equation (\ref{Lag4}). At high mass region where the correction for nesting is negligible (i.e. $C=1$ and $\widetilde C= 0$), the only term contributing significantly to $B_1$ is that with coefficient $\iota$. In addition, $-\lambda$ (and $-\tilde\lambda< -\lambda$) can be neglected in front of unity. Moreover, $H=0$ and $n\pk$ is essentially equal to $N\pk$, so the Lagrangian linear bias reduces to
\beq
B_1(R,\delta)\approx -q J_\delta(R,\delta)=q\frac{\nu-\gamma\,\lav x^2 \rav/\lav x \rav}{\sigma_0\,(1-\gamma^2)}.
\label{simple0}
\eeq
On the contrary, at low masses where the correction for nesting is marked, we can neglect unity in front of $-\lambda$ (or $-\tilde\lambda$) as well as the term with coefficient $\iota$ in front of the remaining ones. These latter roughly balance each other, so $B_1$ is approximately constant (at least within the validity region of the analytic correction for nesting, $\log \nu > -0.4$). 

An approximate analytic expression for any higher order bias parameter can also be readily obtained at high masses from the simple form (\ref{simple0}) of the linear parameter (see Sec.~\ref{LB}).

\section{Analytic Expression of the Correction for Halo Splitting and Grouping}\label{D}

To infer the detailed analytic expression of the peak number density including halo splitting and grouping we must find that of the peak appearance rate (eqs.~[\ref{D0}]-[\ref{aa}])
\beq
r^{a}(R,\delta)= \frac{\partial \ln n\pk(R,\delta)}{\partial \delta}- \frac{\Sigma(R,\delta)}{R}\!\left[\frac{\partial \ln n\pk(R,\delta)}{\partial  \ln R}+\frac{\partial \ln\Sigma(R,\delta)}{\partial  \ln R}\!\right].
\label{aabis0}
\eeq
This can be done accurately. But our aim is just to assess whether spurious halo splitting and grouping, dependent on the specificity of each simulation, may explain the discrepancy between the predicted and empirical halo bias, so we will make a few approximations to simplify the derivation. 

Apart from adopting $M\vir$ masses, we will restrict the derivation to the range of masses covered by simulations (i.e. $\log \nu > 0.1$ or $\log \nu\F > 0$), where there is essentially no peak nesting and we have $n\pk\approx N\pk$. This way we can also approximate the CDM power spectrum by a power-law and adopt the approximation $\lav x^2\rav\approx \lav x\rav^2$ to better than $\sim 3$\%. The latter approximation leads in turn to $\lav x^3\rav\approx \lav x\rav^3$ and $\partial \lav x\rav/\partial\delta\approx \partial \lav x\rav/\partial\ln R\approx 0$. 

In these conditions, $\Sigma(R,\delta)/R\equiv\sigma_0/(\sigma_2 R^2\lav x\rav)$ is approximately equal to $1/\lav x\rav$ and the accretion rate (eq.~[\ref{aabis0}]) becomes
\beq
r^{\rm a}(R,\delta)\approx \frac{\partial \ln N\pk}{\partial\delta}-\frac{1}{\lav x\rav}\left[\frac{\partial \ln N\pk}{\partial \ln R}+1\right]
\eeq
where
\beqa
\frac{\partial \ln N\pk}{\partial \delta} = \frac{-\nu+\gamma\,\lav x^2 \rav(R,\delta)/\lav x \rav(R,\delta)}{\sigma_0\,(1-\gamma^2)}
\approx\frac{-\nu +\gamma\lav x\rav}{\sigma_0(1-\gamma^2)}~~~~~~~~~~~~~~~~~~~~~~~~~~~~~~~~~~~~~~~~~~~~~~~~~~~~~~~~~~~~~~~~~~~~~\\
\frac{\partial \ln N\pk}{\partial \ln R}=1+\left[4-\frac{\chi(R,\delta)+\gamma^2}{1-\gamma^2}\right]\frac{\der\ln (\sigma_2/\sigma_1)}{\der \ln R}+(1-\gamma^2)\left[1+\frac{\chi(R,\delta)+\gamma^2}{1-\gamma^2}+\gamma\nu\frac{\gamma\nu-\gamma^2\lav x\rav}{(1-\gamma^2)^2}\right]\frac{\der\ln \sigma_1}{\der\ln R}
\,~~~~~~\label{J10}\\
\chi(R,\delta)\equiv-\frac{\gamma\nu}{1-\gamma^2} \left[\gamma\nu-(1+\gamma^2)\frac{\lav x^2 \rav}{\lav x \rav}+\gamma^2\frac{\lav x^3 \rav}{\gamma\nu\lav x \rav}\right]\approx -\frac{(\gamma\nu-\gamma\lav x \rav)^2}{1-\gamma^2}+ \frac{1-\gamma}{1+\gamma}\gamma\nu\lav x \rav
\label{Chi0}.\,~~~~~~~~~~~~~~~~~~~~~~~~~~~~
\eeqa

This accretion rate and its derivative
\beq
\frac{\partial r^{\rm a}}{\partial\delta}\approx \left(-\frac{\nu}{\sigma_0}+\frac{\gamma\lav x\rav}{\sigma_0}\right)\left[1+\frac{\gamma^2\delta}{(1-\gamma^2)\lav x\rav}\frac{\der \ln\sigma_1}{\der\ln R}\right]
\eeq
are to be plugged in factors 
\beq
C_{\rm s}(R,\delta)\approx 1+
\frac{1}{2}\Delta_{\rm s}^2(\delta,k_{\rm s})r^{\rm a}(R,\delta)\left[1\!-\!2\frac{(N\pk r^{\rm a})_{\rm s}(R_{\rm s},\delta)}{(N\pk r^{\rm a})_{\rm s}(R,\delta)}\right]~~~~~~
C_{\rm g}(R,\delta)\approx 1+
\frac{1}{2}\Delta_{\rm g}^2(\delta,k_{\rm g})r^{\rm a}(R'_{\rm s},\delta)\left[2\!-\!\frac{(N\pk r^{\rm a})_{\rm g}(R_{\rm g},\delta)}{(N\pk r^{\rm a})_{\rm g}(R,\delta)}\right],~~~\nonumber
\label{averbisE0}
\eeq
(see eqs.~[\ref{nk}]-[\ref{D0}]) and [\ref{noavertris}]-[\ref{D0bis}], respectively), where $(N\pk r^{\rm a})_{\rm s}(R,\delta)$ stands for $N\pk(R,\delta) r^{\rm a}(R,\delta)$ and $(N\pk r^{\rm a})_{\rm g}(R,\delta)$ stands for $N\pk(R,\delta)r^{\rm a}(R'_{\rm s},\delta)$, and in their $\delta$-derivatives, equal to leading order in $\Delta_{\rm s}$ and $\Delta_{\rm g}$ to
\beqa
\frac{\partial \ln C_{\rm s}}{\partial \delta}
\!\approx\!- \frac{\Delta^2_{\rm s}(\delta,k_{\rm s})}{2}r^{\rm a}(R,\delta)Q_{\rm s}\left\{\!2\frac{\partial \ln\Delta_{\rm s}}{\partial \delta}\!+\!\frac{\partial \ln r^{\rm a}(R,\delta)}{\partial \delta}\!+\!\frac{Q_{\rm s}\!-\!1}{Q_{\rm s}}
\!\!\left[\frac{\partial \ln(N\pk r^{\rm a})_{\rm s}(R_{\rm s},\delta)}{\partial \delta}\!-\!\frac{\partial \ln(N\pk r^{\rm a})_{\rm s}(R,\delta)}{\partial \delta}\right]\!\right\}
\label{part10}~~~~~~~~~~~~\\
\frac{\partial \ln C_{\rm g}}{\partial \delta}
\!\approx \!-\frac{\Delta^2_{\rm g}(\delta,k_{\rm g})}{2}r^{\rm a}(R'_{\rm s},\delta)Q_{\rm g}\!\left\{\!2\frac{\partial \ln\Delta_{\rm g}}{\partial \delta}\!+\!\frac{\partial \ln r^{\rm a}(R'_{\rm s},\delta)}{\partial \delta}\!+\!\frac{Q_{\rm g}\!-\!1}{Q_{\rm g}}\!\!
\left[\frac{\partial \ln(N\pk r^{\rm a})_{\rm g}(R_{\rm g},\delta)}{\partial \delta}\!-\!\frac{\partial \ln(N\pk r^{\rm a})_{\rm g}(R,\delta)}{\partial \delta}\right]\!\right\},~~~~~~~
\label{part1E0}
\eeqa
being
\beq
Q_{\rm s}(R,\delta)\equiv 1-2\frac{(N\pk r^{\rm a})_{\rm s}(R_{\rm s},\delta)}{(N\pk r^{\rm a})_{\rm s}(R,\delta)}~~{\rm and}~~Q_{\rm g}(R,\delta)\equiv 2-\frac{(N\pk r^{\rm a})_{\rm g}(R_{\rm g},\delta)}{(N\pk r^{\rm a})_{\rm g}(R,\delta)}.\nonumber
\eeq
Other quantities appearing in the right-hand-members of equations (\ref{part10}) and (\ref{part1E0}) are
\beqa
\frac{\partial \Delta_{\rm s}}{\partial \delta}=\frac{\der \delta(t_{\rm s})}{\der t_{\rm s}}\frac{\der t_{\rm s}}{\der \delta}-1
\approx \frac{\delta(t_{\rm s})}{\dot \delta}\!\left[1-k_{\rm s}\tau\cc(t_0)\frac{3}{2}\frac{\dot a}{a}\right]\left[d{\cal D}(t_{\rm s})\frac{\dot a(t_{\rm s})}{a(t_{\rm s})}+\frac{\der\ln(\delta_c^{\rm th}/ D^2)}{\der t_{\rm s}}\right]-1~~~~~~~~~~~~~~~~~~~~~~~~~~~~~~~~~\\
\frac{\partial \Delta_{\rm g}}{\partial \delta}=1-\frac{\der \delta(t_{\rm g})}{\der t_{\rm g}}\frac{\der t_{\rm g}}{\der \delta}
\approx 1-\frac{\delta(t_{\rm g})}{\dot \delta}\!\left[1+k_{\rm g}\tau\cc(t_0)\frac{3}{2}\frac{\dot a}{a}\right]\left[d{\cal D}(t_{\rm g})\frac{\dot a(t_{\rm g})}{a(t_{\rm g})}+\frac{\der\ln(\delta_c^{\rm th}/ D^2)}{\der t_{\rm g}}\right],~~~~~~~~~~~~~~~~~~~~~~~~~~~~~
\label{lastD0}
\eeqa
where a dot denotes time-derivative at $t$, $t_{\rm s}$ and $t_{\rm g}$ stand for $t-k_{\rm s}\tau\cc(t)$ and $t+k_{\rm g}\tau\cc(t)$, respectively, and $t_0$ is the present cosmic time.

\end{appendix}

\end{document}